\documentclass[10pt,conference]{IEEEtran}
\IEEEoverridecommandlockouts
\usepackage{cite}
\usepackage{amsmath,amssymb,amsfonts}
\usepackage{algorithmic}
\usepackage{graphicx}
\usepackage{textcomp}
\usepackage{xcolor}
\usepackage{amsmath}
\usepackage{multirow}
\usepackage{algorithm}
\usepackage[hypcap=false]{caption}
\usepackage{subcaption}
\usepackage[capitalise]{cleveref}
\usepackage[font=small,skip=5pt]{caption}

\DeclareCaptionFont{mysize}{\fontsize{8}{9.6}\selectfont}
\def\BibTeX{{\rm B\kern-.05em{\sc i\kern-.025em b}\kern-.08em
    T\kern-.1667em\lower.7ex\hbox{E}\kern-.125emX}}
\begin{document}

\title{iMove: Exploring Bio-impedance Sensing for Fitness Activity Recognition
}


\author{\IEEEauthorblockN{
        Mengxi Liu\IEEEauthorrefmark{1}\IEEEauthorrefmark{3},
        Vitor Fortes Rey\IEEEauthorrefmark{1}\IEEEauthorrefmark{2}\IEEEauthorrefmark{3},
        Yu Zhang\IEEEauthorrefmark{1}\IEEEauthorrefmark{2}, \\
        Lala Shakti Swarup Ray\IEEEauthorrefmark{1},
        Bo Zhou\IEEEauthorrefmark{1}\IEEEauthorrefmark{2},and
        Paul Lukowicz\IEEEauthorrefmark{1}\IEEEauthorrefmark{2}
        }
    \IEEEauthorblockA{\IEEEauthorrefmark{1}German Research Center for Artificial Intelligence (DFKI), Kaiserslautern, Germany}
    \IEEEauthorblockA{\IEEEauthorrefmark{2}Department of Computer Science, RPTU Kaiserslautern-Landau, Kaiserslautern, Germany}
    \IEEEauthorblockA{Email: firstname.lastname@dfki.de}
    \IEEEauthorblockA{\IEEEauthorrefmark{3} equal contribution}
    
}

\IEEEpeerreviewmaketitle
\setcounter{page}{1}
\maketitle

\begin{abstract}
Automatic and precise fitness activity recognition can be beneficial in aspects from promoting a healthy lifestyle to personalized preventative healthcare. While IMUs are currently the prominent fitness tracking modality, through iMove, we show bio-impedence can help improve IMU-based fitness tracking through sensor fusion and contrastive learning.To evaluate our methods, we conducted an experiment including six upper body fitness activities performed by ten subjects over five days to collect synchronized data from bio-impedance across two wrists and IMU on the left wrist.The contrastive learning framework uses the two modalities to train a better IMU-only classification model, where bio-impedance is only required at the training phase, by which the average Macro F1 score with the input of a single IMU was improved by 3.22 \% reaching 84.71 \% compared to the 81.49 \% of the IMU baseline model. We have also shown how bio-impedance can improve human activity recognition (HAR) directly through sensor fusion, reaching an average Macro F1 score of 89.57 \%  (two modalities required for both training and inference) even if Bio-impedance alone has an average macro F1 score of 75.36 \%, which is outperformed by IMU alone. In addition, similar results were obtained in an extended study on lower body fitness activity classification, demonstrating the generalisability of our approach.Our findings underscore the potential of sensor fusion and contrastive learning as valuable tools for advancing fitness activity recognition, with bio-impedance playing a pivotal role in augmenting the capabilities of IMU-based systems.

\end{abstract}

\begin{IEEEkeywords}
bio-impedance sensing, human activity recognition, contrastive learning, sensor fusion
\end{IEEEkeywords}

\section{Introduction}


Regular physical activity is crucial for both mental and physical well-being \cite{dishman1985determinants,penedo2005exercise}. 
Fitness activity, a planned and repetitive subset of physical activity \cite{fu2020unconstrained}, plays an important role in personalized preventative healthcare. 
Automating the tracking of fitness activities can provide effective feedback, motivating users to exercise correctly and regularly. 
The field of Human Activity Recognition (HAR) has evolved from manual or semi-automated logging to advanced methods such as smart ambient objects \cite{fu2020unconstrained, guo2018device} and wearable devices \cite{zhou2016never,bian2019passive,qi2018hybrid}. 
These solutions aim to be precise, convenient, unobtrusive, and user-friendly, promoting a sustainable healthy lifestyle.

Computer vision techniques are widely used for fitness activity recognition, showing high performance with advanced deep learning methods \cite{rangari2022video, aiman2023video}. However, these methods raise privacy concerns, limiting their use in private settings. Sensor-based solutions offer an alternative, employing various sensing principles like kinematic sensing (e.g., IMU), field-based sensing (e.g., Capacitive), wave-based sensing (e.g., WIFI), and physiological sensing (e.g., Electrophysiological) \cite{bian2022state}. IMUs, integrated into devices like smartwatches, are popular due to their low power consumption and ease of deployment. 
Electrophysiological sensors, like EMG, provide precision tracking but require placement on specific body parts \cite{karolus2021facilitating}. This limitation can be addressed by passive capacitive sensing methods \cite{bian2019passive}, but they often suffer from low robustness due to their high sensitivity to the environment.


Bio-impedance sensing, a type of electrophysiological signal, monitors conductive information between two electrodes, enabling monitoring of multiple body parts (e.g., hand-to-hand), unlike IMU sensors. It is dependent upon body status, making it more robust than capacitive sensing. 
While existing bio-impedance solutions mainly focus on physiological indicators and gestures \cite{piuzzi2018low,xu2016wrist,zhang2015tomo,chen2021bio}, its application in fitness activity recognition is not widely explored in current literature.
In addition, contrastive learning has demonstrated an improvement over supervised and unsupervised learning in Human Activity Recognition for health \cite{tang2020exploring}, but contrastive learning between different modalities is not widely studied.

In this work, we present iMove, our solution including hardware and machine learning methods to explore how bio-impedance sensing can improve wrist-worn wearables for fitness activity recognition. 
To evaluate the performance of iMove for fitness activity recognition, we conducted an experiment including six fitness activities performed by ten subjects over five days.
In addition, we propose an algorithm based on contrastive learning to train better IMU-only classification models using bio-impedance sensing that is only required at the training time.

Overall, in this paper, we present the following contributions (all reported percentages are average Macro F1 scores with leave-one-person-out cross-validation method):

\begin{enumerate}
    \item Through a novel contrastive training framework, with shared latent representation from different modalities, we used bio-impedance to enhance a mono-modality (IMU) model, where bio-impedance is only required at the training phase. Our method increased the average Macro F1 score by 3.22 \% for upper body activities and 2.79 \% for lower body ones.

    \item Through sensor fusion, where the neural network model takes both the bio-impedance and IMU signals as input, two modalities complement each other and significantly improve the recognition result, obtaining 89.57\% for upper body activities and 81.74\% for lower body ones.
    
    \item We have also demonstrated how single-channel bio-impedance sensing across two wrists is a viable stand-alone modality for fitness activity recognition for both upper body activities (75.36 \%) and lower body ones (70.02 \%), albeit inferior to a single IMU sensor on one wrist (81.49 \% and 77.99 \%).
    
\end{enumerate}

\section{Related work}

\subsection{Bio-impedance sensing-based wearables for HAR}
Bio-impedance methods involve the introduction of an alternating electric current of very low intensity into a biological medium (such as the human body, tissue, or cell culture), leading to a voltage drop, the magnitude of which is proportional to the electrical impedance of the biological medium \cite{roa2013applications}. 
Thus, many existing works based on bio-impedance sensing wearables for HAR tasks have been demonstrated in the past decades in many application scenarios, like human physiological index monitoring \cite{xu2016wrist,sel2020wrist }, gesture recognition \cite{zhang2015tomo, ma2023two}, automatic dietary monitoring \cite{zhang2016generic}, and sport science\cite{ackland2012current}. 
For instance, the bio-impedance based heartbeat monitoring system with four electrodes in the wristband configuration was first proposed in work \cite{xu2016wrist}, whose result showed that impedance variation at the wrist closely matched with the heartbeat signal acquired from a standard piezoelectric finger pulse transducer.
Kaan et al \cite{sel2020wrist}. proposed a wrist-worn respiration monitoring device based on bio-impedance to estimate respiration parameters using gold e-tattoos.
Besides, the work \cite{zhang2015tomo} showed a gesture recognition solution based on bio-impedance tomography using multiple electrodes with a near to perfect accuracy for eleven gestures by multiple pairs of electrodes in real-time. 
Ma et al \cite{ma2023two}. proposed a frequency-scan system for gesture recognition using the bio-impedance information, which only required two electrodes, which also achieved a very high recognition accuracy for four common gestures and a group of six pinch gestures.
The bio-impedance based method was also widely applied in other scenarios.
For example, Zhang et al \cite{zhang2016generic}. proposed a generic sensor fabric for swallowing sensing based on resistive pressure and bio-impedance sensors, which were integrated into a shirt collar; the result conﬁrmed the signal performance of both sensor types for swallowing spotting. 
The bio-impedance sensing-based solutions for biometric studies have been proposed in many works \cite{van2020portable, holz2015biometric}.
Although the bio-impedance sensing based solution is also widely used in the sports science area \cite{ackland2012current}, like assessing knee joint health \cite{hersek2016wearable} and muscle fatigue \cite{vescio2012muscle}, most works mainly focused on a physiologic monitoring and specific part of the body, like hands (gesture recognition), leg (knee joint health) or specific muscles. 
Compared to the existing application scenarios, bio-impedance sensing based full body activity recognition has not been widely studied according to contemporary literature.

\subsection{Sensor fusion in neural networks for HAR}
The concurrent use of multiple sensors for human activity recognition provides a rich data source for complex activity recognition and improvement of recognition accuracy \cite{aguileta2019multi}.
Thus, sensor fusion methods have been widely studied in the contemporary literature. In \cite{munzner2017cnn} authors explored sensor fusion for HAR using IMUs, comparing early, late and hybrid fusion approaches, among others. Other approaches such as \cite{choi2018confidence} learn how to weight the contribution of the different sensors. Other works, such as \cite{liu2020giobalfusion, zhou2022tinyhar} use attention modules for multi-sensor information fusion.

On the other hand, the aforementioned approaches, as in most of the sensor fusion solutions, require all sensor modalities to be available both in the training and inference phase \cite{zhou2022tinyhar,gravina2017multi}.
However, there are two obvious disadvantages of these solutions: the input complexity of the model could be increased along with the modality number; using multiple sensing modalities could be difficult to realize in typical daily life environments. Those disavantages are addressed by currently proposed methods, which only require all sensor modalities available at training time \cite{fortes2022learning, lago2021using, yang2022more} while fewer sensor modalities are required at the inference time.
For example, the work More to Less \cite{yang2022more} transfers knowledge between different models with different input modalities only when the performance of one model is better than the other, with only one modality required at the inference time. Still, the transfer requires labeled data. The work \cite{lago2021using} improves single sensor performance using clustering and boosting in the training phase and supports unlabelled data.
The work Learning from the Best \cite{fortes2022learning} proposed a method based on contrastive learning which can transfer the knowledge between each model bidirectionally and also supports unlabelled data. But that work did not present transferring knowledge between two different modalities, as the method focuses on sensors of the same modality (IMU) located in different parts of the body.

Inspired by the work Learning from the Best, we proposed a contrastive learning based method that transfers knowledge between different modalities, such as bio-impedance sensor to IMU sensor, to improve the performance of a single IMU sensor for the full-body fitness activity recognition task.
Besides, we also proposed a hyperparameter to control the knowledge transmission amount and direction in this work.
\section{Sensing prototype implementation}

\begin{figure}[!t]
\includegraphics[width=1.0\linewidth]{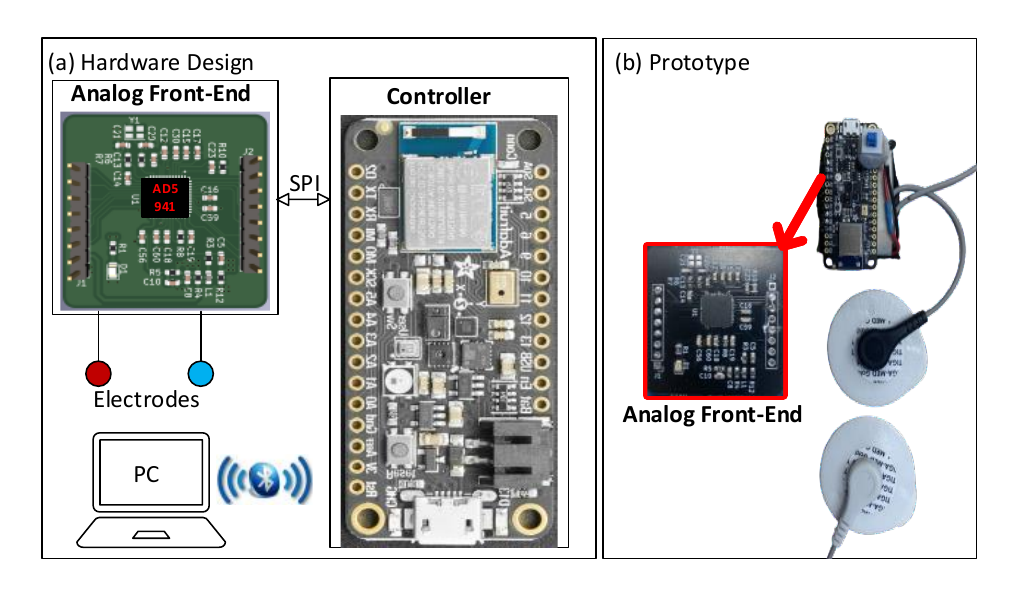}
\caption{Hardware design and prototype of iMove}
\label{fig: hardware_design}
\end{figure}

Bio-impedance measurements are based on the fact that biological mediums act as either conductors, dielectrics, or insulators of electrical current, depending on their composition. 
This leads to a frequency-dependent bio-impedance, which can be used to gain insight into the physiology and pathology of tissues and cells \cite{naranjo2019fundamentals}.
The iMove wearable hardware prototype shown in \cref{fig: hardware_design} was designed to monitor the bio-impedance signal variation that occurs as users perform fitness activity, allowing for precise tracking of them.
The iMove prototype consists of four modules: the analog front-end (AFE) module, the control module, the electrodes, and the power supply.
The chip AD5941 (Analog Devices) was selected as the core component of the analog front-end module of iMove, as it can both generate the voltage stimuli as sinuous signals with a configurable frequency from 0.015 Hz to 200 kHz and measure the response current signal with an integrated high-speed trans-impedance amplifier.
It also has an integrated FFT hardware accelerator to decipher the real and imaginary components from the measurement.
An nRF52840 (Nordic) from Feather BLE Sense is selected as the controller module that drives the AFE through an SPI bus and transmits the measurement result to the computer via Bluetooth. 
Besides, there is an IMU sensor on the Feather BLE Sense board.
The wet Ag/AgCI electrodes interface the user's body with the AFE.
The work current is around 20 mA.
A compact lithium battery with a 500 mAh capacity powers the system, which could power the system for around one day. 

\section{Experiment setup}
Ten volunteers (4 females and 6 males) aged from 21 to 32 years old with weights from 50.6 kg to 85.15 kg were invited to take part in an experiment to assess the performance of iMove with bio-impedance sensing for recognizing fitness activities. 
The participants were asked to do six different fitness activities five times within five different days, each time lasting approximately 15 minutes while wearing iMove indoors.
The selected six fitness activities come from Pamela \footnote{https://www.youtube.com/@PamelaRf1}, including the \textit{Box}, \textit{Biceps curl}, \textit{Chest press}, \textit{Shoulder and chest press}, \textit{Arm hold and shoulder press}, and \textit{Arm opener} as shown in \cref{fig: raw signal}. 
During the experiment, both bio-impedance and IMU data were collected, and the data was transmitted to a web application on the laptop via Bluetooth. 
The web application was written in JavaScript, which was used to observe the bio-impedance and IMU signal in real-time and store the data in text files on the laptop.
A front-facing camera was used to record participants with the video used to label the data after the experiment.
The TRAINSET tool from Geocene was used to label the time series data.
The sampling rate of bio-impedance data and IMU is 20 Hz after synchronization.
The instances with a window size of 50 were generated by a sliding window approach with a slide step of 10.

\subsection{Bio-impendance sensing configuration}
\begin{figure}[!t]
\includegraphics[width=1.0\linewidth ]{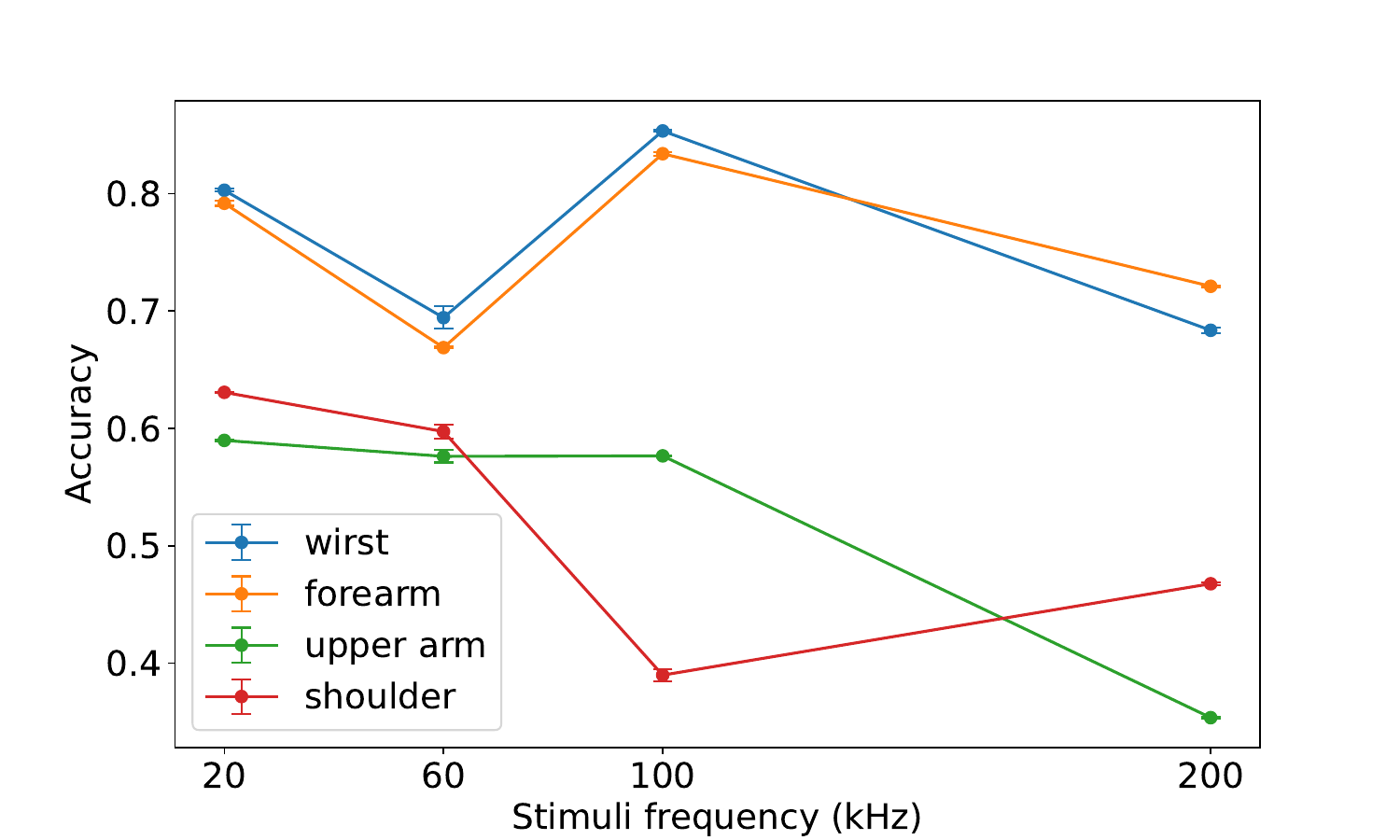}
\caption{Upper body classification results with different experiment configurations (stimuli frequency and electrode location)}
\label{fig: frequency_selection}
\end{figure}

The stimuli frequency used for the bio-impedance measurement has a great effect on the relationship of the measurement to the applied motion and its relationship with the resulting motion artifact \cite{comert2014impedance}.  
In addition, the location of electrodes also has an effect on the bio-impedance measured result.
To obtain high-quality bio-impedance data, a primary experiment was conducted to select a better experiment configuration, where we tested four different stimuli frequencies, such as 20 kHz, 60, kHz, 100 kHz, and 200 kHz, and four different electrode locations, such as wrist, forearm, upper arm, and shoulder.
Then we used a simple KNN classifiers to process the measured bio-impedance data; the classification result is shown in \cref{fig: frequency_selection}, from which we can find that the configuration of stimuli frequency with 100 kHz and electrodes on the wrist achieved the highest classification accuracy.
The result of the sensor location was in line with our expectations, as the sensing range is the largest when attaching electrodes to wrists.
Thus, this configuration with 100 kHz stimuli and a wrist-worn electrode was selected for our experiments.

\subsection{Data Collection}

\begin{figure*}[h]
\footnotesize
\centering
\includegraphics[width=1.0\linewidth]{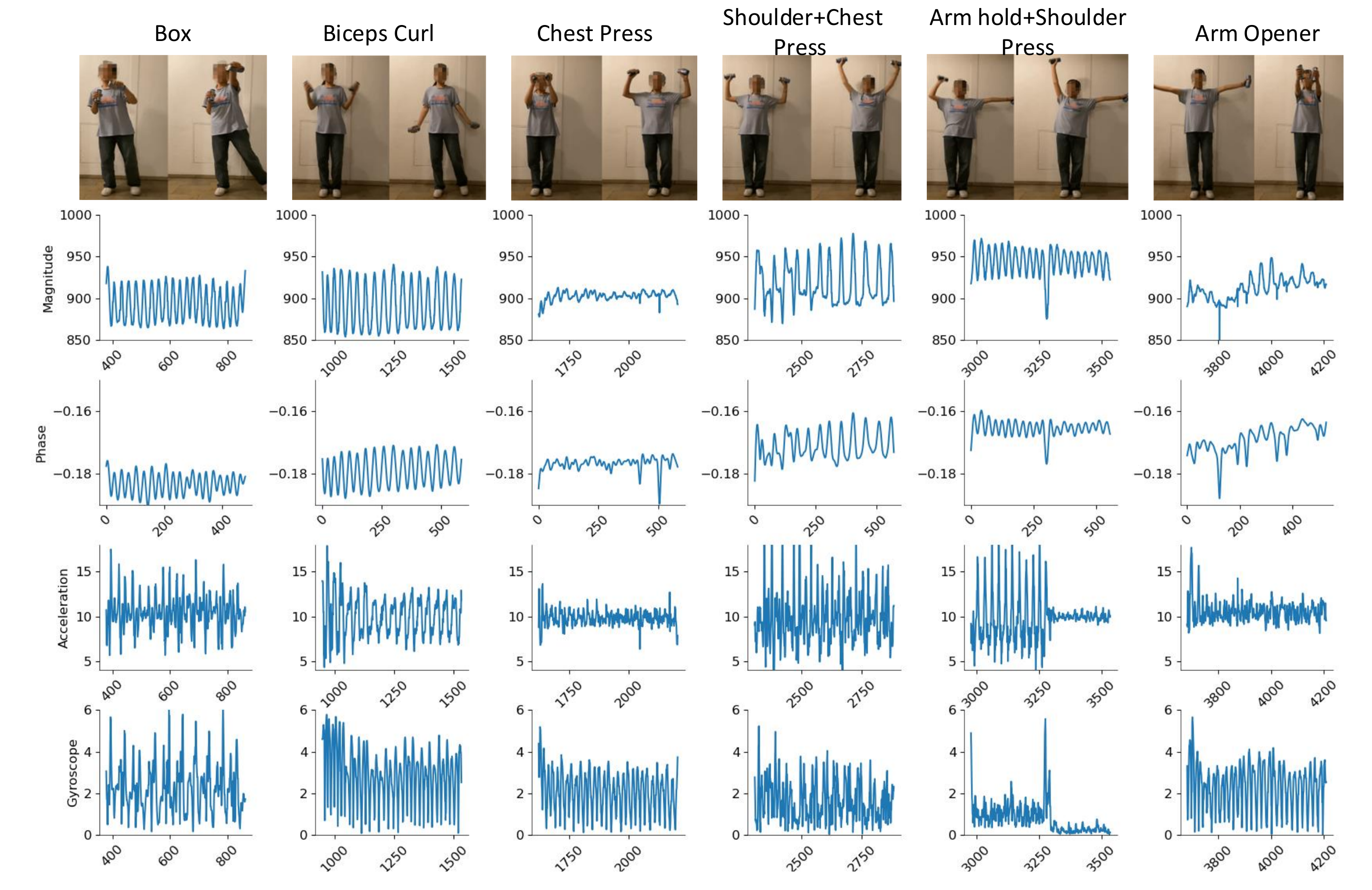}
\caption{Bio-impedance and IMU signal of six upper body fitness activities (The IMU signal is the L2-Norm of raw data. \textbf{Chest Press}: participant holds a one-kilogram dumbbell in each hand and presses the dumbbells away from the chest until arms are fully extended, and brings them back to the starting position; \textbf{Shoulder+Chest Press}: participant adds an action of lifting the dumbbells  from shoulder height to above his/her head when his/her arms are fully extended based on chest press 1 activity; \textbf{Arm hold+shoulder press}: participant holds a dumbbell with an overhand grip at shoulder height by one hand, push the dumbbell overhead by fully extending his/her arm, while keeping the other arm at shoulder height, change the hand to hold the dumbbell after more than ten times)}
\label{fig: raw signal}
\end{figure*}

In the experiment, both bio-impedance and IMU data (accelerometer and gyroscope) were collected to evaluate the performance of bio-impedance sensing modality.
\cref{fig: raw signal} shows the raw signal from the bio-impedance sensor and IMU while performing different fitness activities. 
The bio-impedance data includes two channels: magnitude and phase; The original IMU signal includes three-channel acceleration and three-channel angular speed. 
We use the norm of the accelerometer ($\sqrt{ax^2+ay^2+az^2}$) and gyroscope ($\sqrt{gx^2+gy^2+gz^2}$) instead of the raw values, as this provided better baseline HAR results when compared to the raw channels. This is most likely due to the reduction of the effect of IMU sensor orientation. Moreover, using only the norm also reduces the model size.

It can be observed that fitness activities can lead to distinguishing variations of bio-impedance signal, especially in activities where arm bending movement is inclusive, such as \textit{Box} and\textit{ Biceps Curl}. 
However, activities, like \textit{Chest press} and \textit{Arm opener}, where the majority of the movement comes from the shoulder joint, cause smaller bio-impedance variation compared to bending the arm.
Besides, the repetitive peaks in bio-impedance come from the repetitive activity, which looks more obvious than the IMU signal. It's worth mentioning that one bio-impedance sensor can monitor the movement of both arms; for example, the subjects move one arm while keeping another hand still when performing the activity \textit{Arm hold and shoulder press}, whichever arm moves, it can cause significant bio-impedance signal changes, while IMU sensor can only track the movement of the arm where the IMU sensor is worn.

\begin{figure*}[ht]
\footnotesize
\centering
     \begin{subfigure}[b]{0.33\textwidth}
         \centering
         \includegraphics[width=\textwidth]{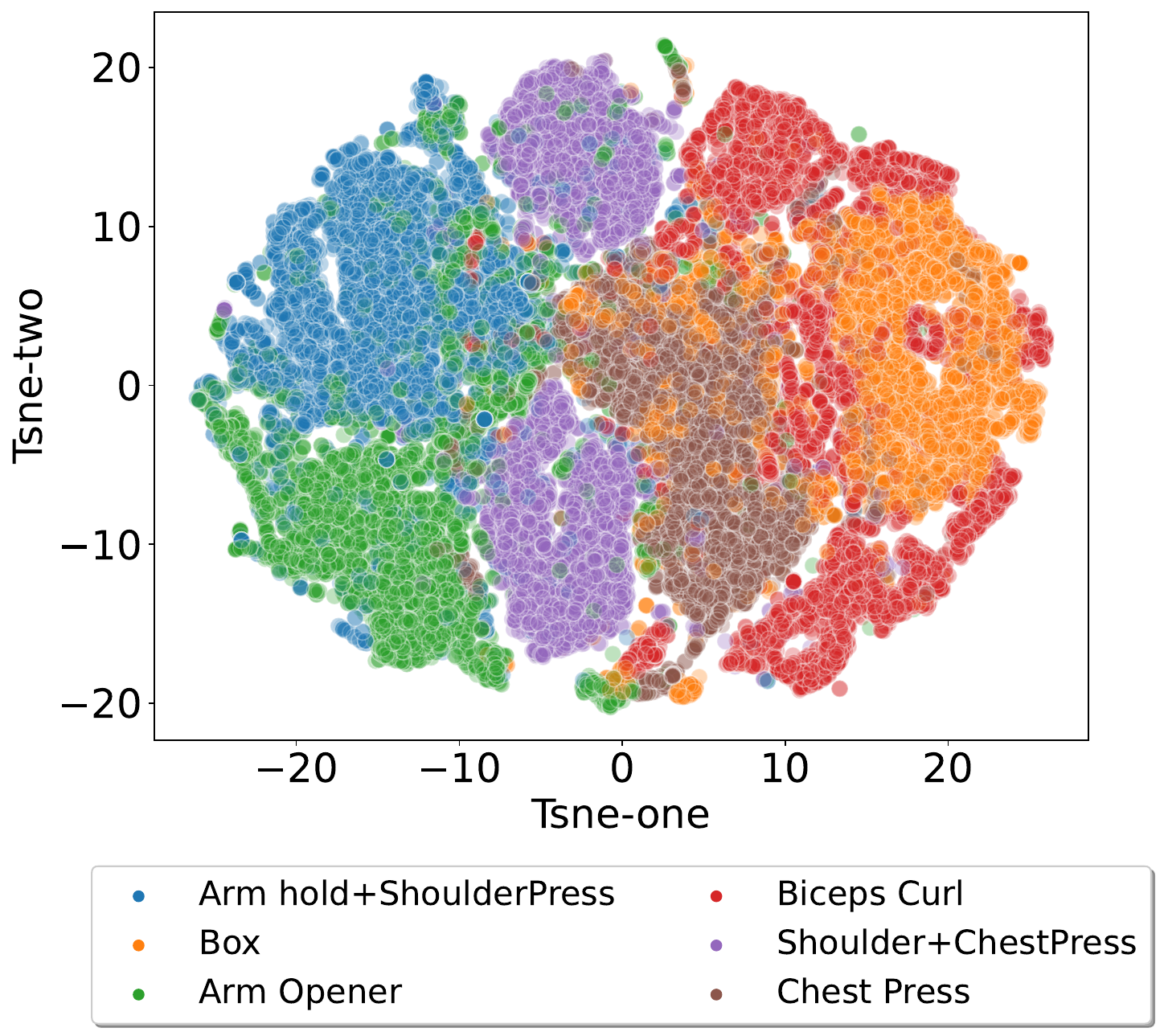}
         \caption{Bio-impedance}
         \label{fig:tsne_mp_50}
     \end{subfigure}
     \hfill
     \begin{subfigure}[b]{0.33\textwidth}
         \centering
         \includegraphics[width=\textwidth]{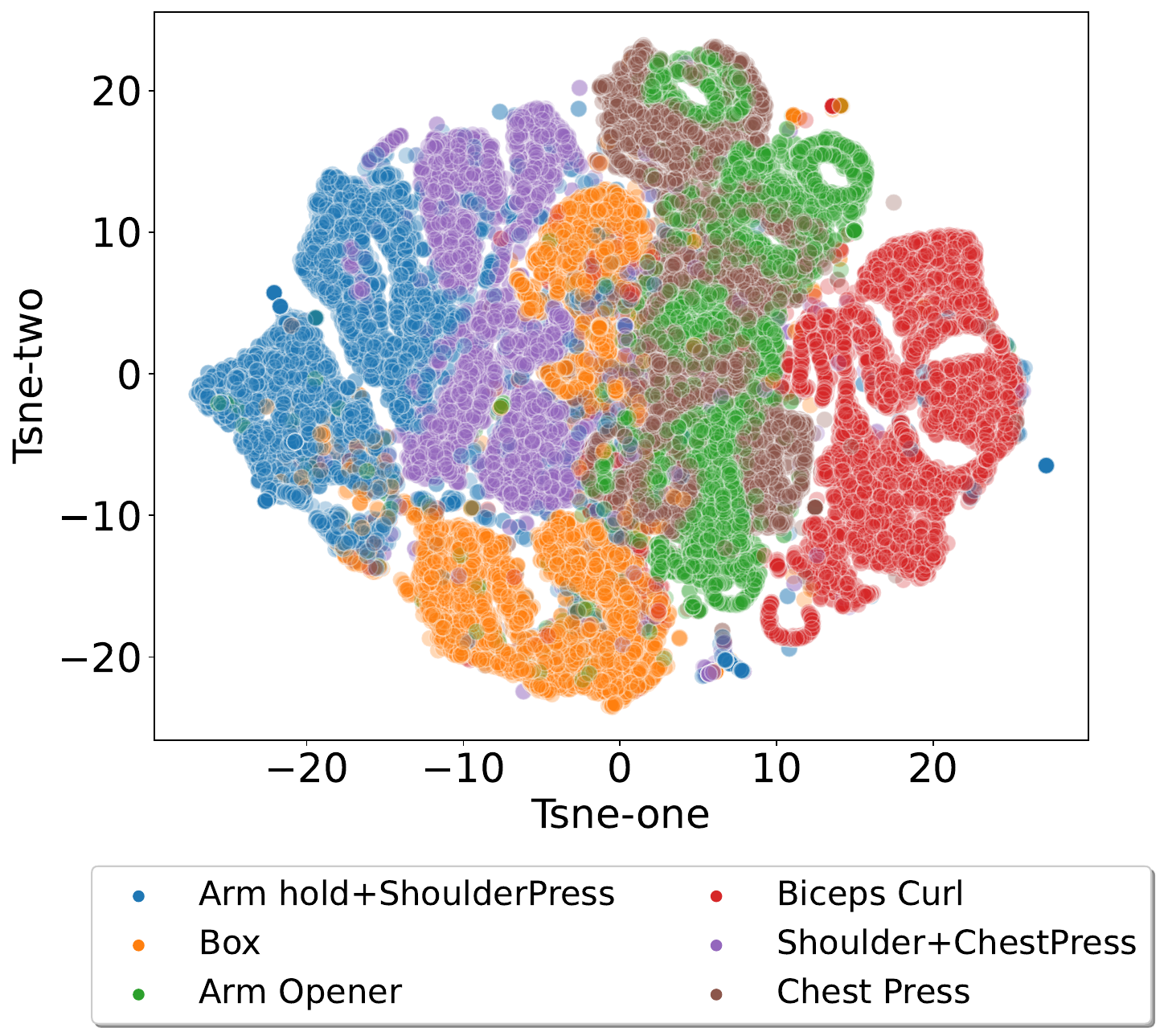}
         \caption{IMU}
         \label{fig:tsne_ag_50}
     \end{subfigure}
     \hfill
     \begin{subfigure}[b]{0.33\textwidth}
         \centering
         \includegraphics[width=\textwidth]{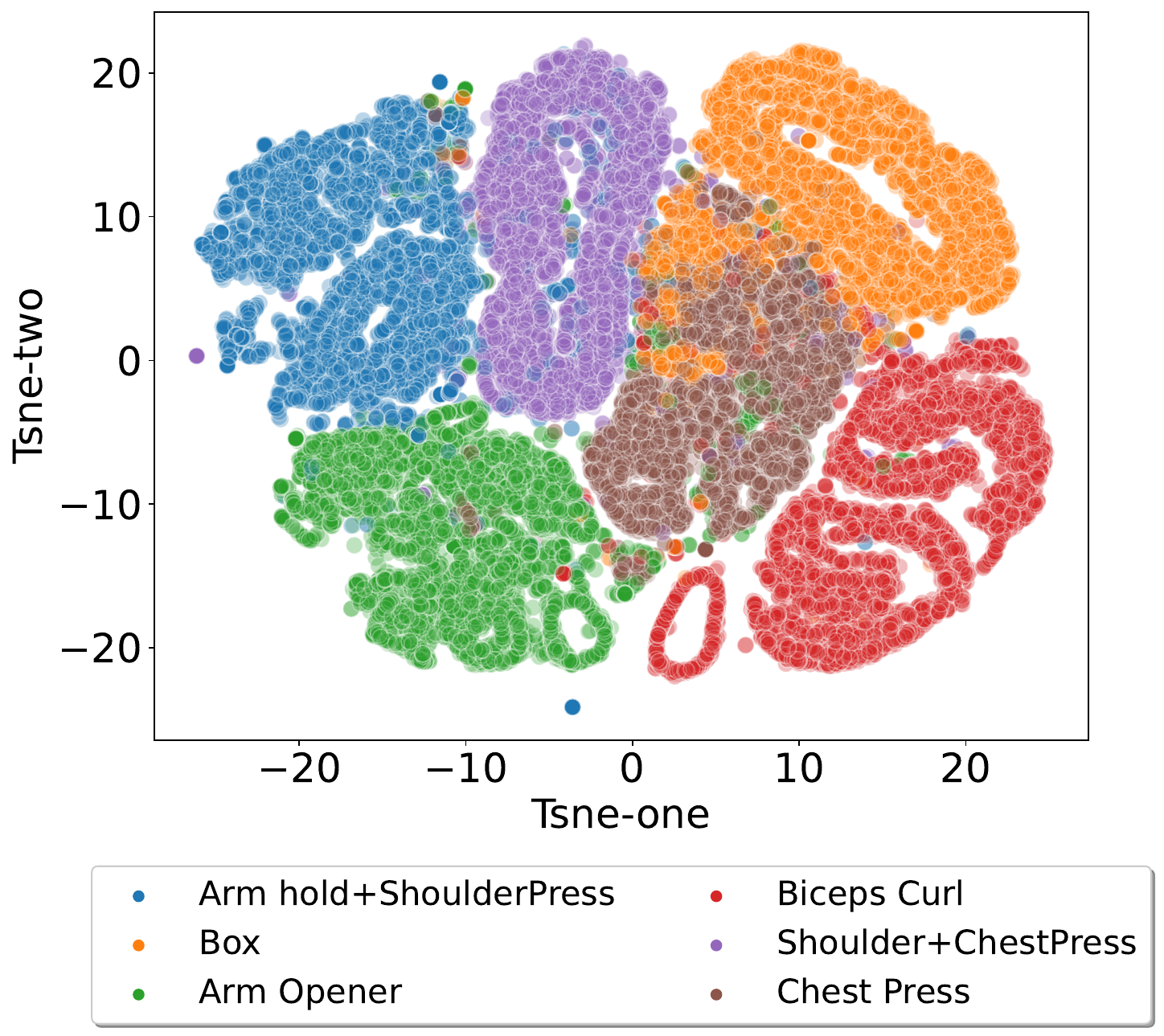}
         \caption{Bio-impedance and IMU}
         \label{fig:tsne_agmp_50}
     \end{subfigure}
    \caption{t-SNE plots of six upper body fitness activities with different sensing modalities}
    \label{fig:t-sne}
\end{figure*}

\cref{fig:t-sne} shows the t-SNE plots of six fitness activities with different sensing modalities. 
We employed Principal Component Analysis (PCA) to extract the 20 most important components from the raw instances with a window size of 50.
Subsequently, we utilized t-SNE to further reduce these 20 components into the two most significant ones.
Both the PCA and TSNE functions come from the Python package \textit{scikit-learn}. 
It can be observed that the clusters of these activities, like \textit{Box} and \textit{Chest Press}, tend to mix when the features from bio-impedance data as shown in \cref{fig:tsne_mp_50}.
However, the clusters of these activities, like \textit{Arm opener} and \textit{Chest press}, tend to be intertwined when the features from IMU data are shown in \cref{fig:tsne_ag_50}.
The boundary of the clusters of each activity becomes much clearer when the features from both bio-impedance and IMU data are provided as shown in \cref{fig:tsne_agmp_50}.
The result from the t-SNE plot demonstrates that the concurrent use of the two sensing modalities has great potential to improve the classification result of the six different activities.

\section{Fitness HAR using each modality and their fusion}
\begin{figure*}[!t]
\footnotesize
\centering
\includegraphics[width=1.0\linewidth]{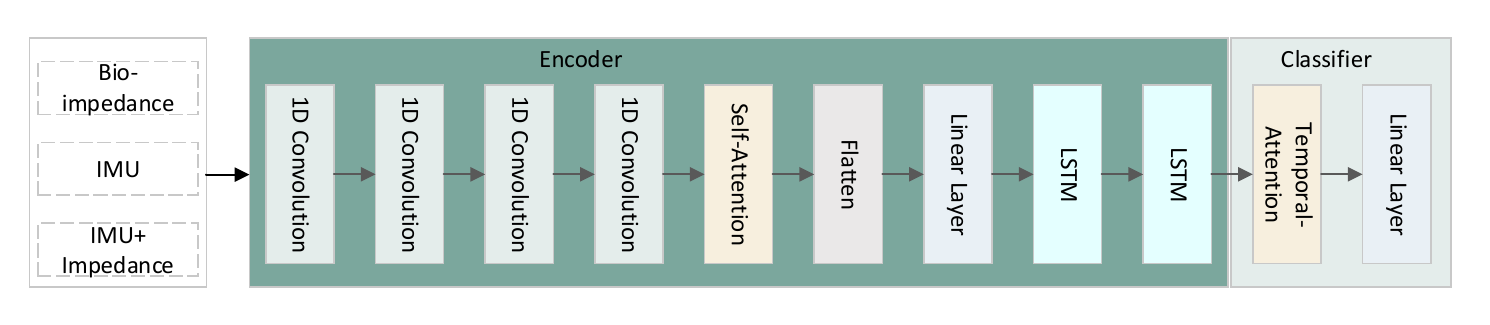}
\caption{Backbone architecture of the neural network used for classification (Backbone)}
\label{fig: architecture neural network}
\end{figure*}

\cref{fig: architecture neural network} shows the Backbone neural network architecture for fitness activity recognition, including an encoder for feature extraction and a classifier.
The Backbone model was built upon the framework from TinyHAR \cite{zhou2022tinyhar}, as it demonstrated outstanding performance on several HAR datasets with hundred-kilobyte model sizes.
In the encoder, the raw sensor data was processed by four one-dimensional channel-wise convolutions, a self-attention module, one linear layer, and two LSTM layers sequentially to extract our $(10\times 40)$ latent representation, which is followed by the classifier to output the recognition result, including a temporal attention module, and one linear layer.
The cross-entropy loss function and Adam optimizer with a learning rate of 5e-4 and 1024 batch size were selected to train this model.
In addition, instances were weighted more for rarer classes using $\frac{\max{(M)}}{m_i}$ where the distribution of the number of data samples per class for the training data is given $M=[m_1, ... , m_{n_a}]$.
The model was trained for 100 epochs with an early stopping using the patience of 30 to avoid overfitting. The validation dataset is the one subject selected randomly from the training dataset. 

To evaluate the performance of the bio-impedance sensing modality comprehensively, the bio-impedance, IMU data, and the combination of bio-impedance and IMU data were evaluated as model inputs separately. 
The leave-one-subject-out cross-validation method was employed in this work to demonstrate the model can recognize the activity across subjects. 
Each subject from the ten subjects was selected as the test subject iteratively; the other nine subjects were selected as train and validation datasets. 

\begin{figure*}[!t]
\footnotesize
\centering
     \begin{subfigure}[b]{0.33\textwidth}
         \centering
         \includegraphics[width=\textwidth]{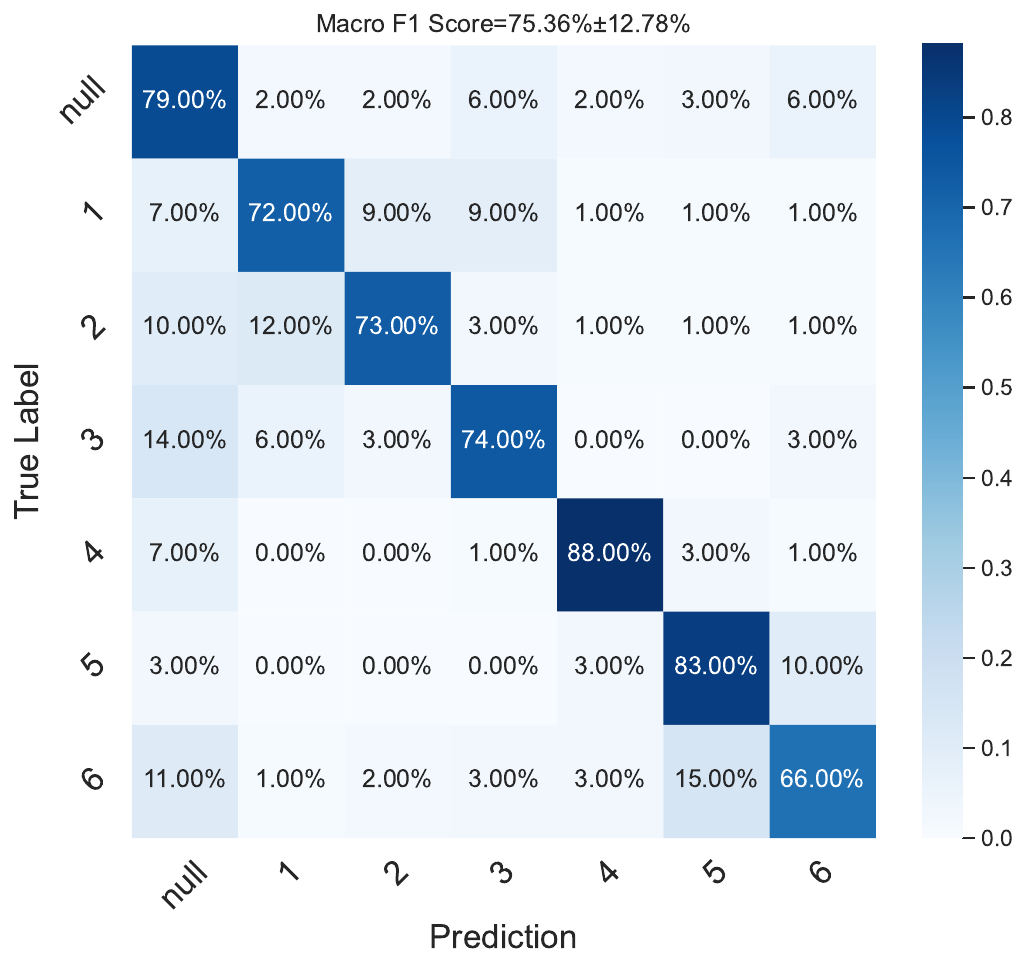}
         \caption{Bio-impedance}
         \label{fig:confusion_matrix_imp}
     \end{subfigure}
     \hfill
     \begin{subfigure}[b]{0.33\textwidth}
         \centering
         \includegraphics[width=\textwidth]{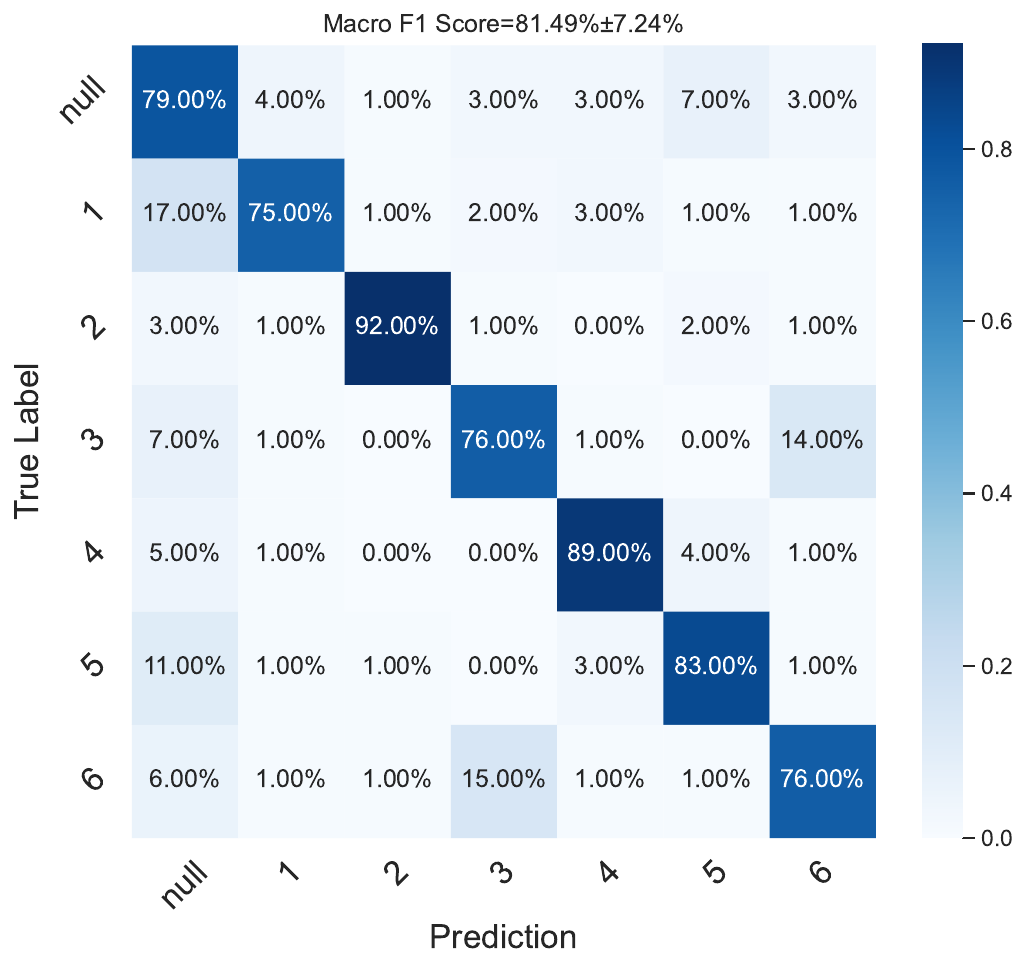}
         \caption{IMU}
         \label{fig:confusion_matrix_imu}
     \end{subfigure}
     \hfill
     \begin{subfigure}[b]{0.33\textwidth}
         \centering
         \includegraphics[width=\textwidth]{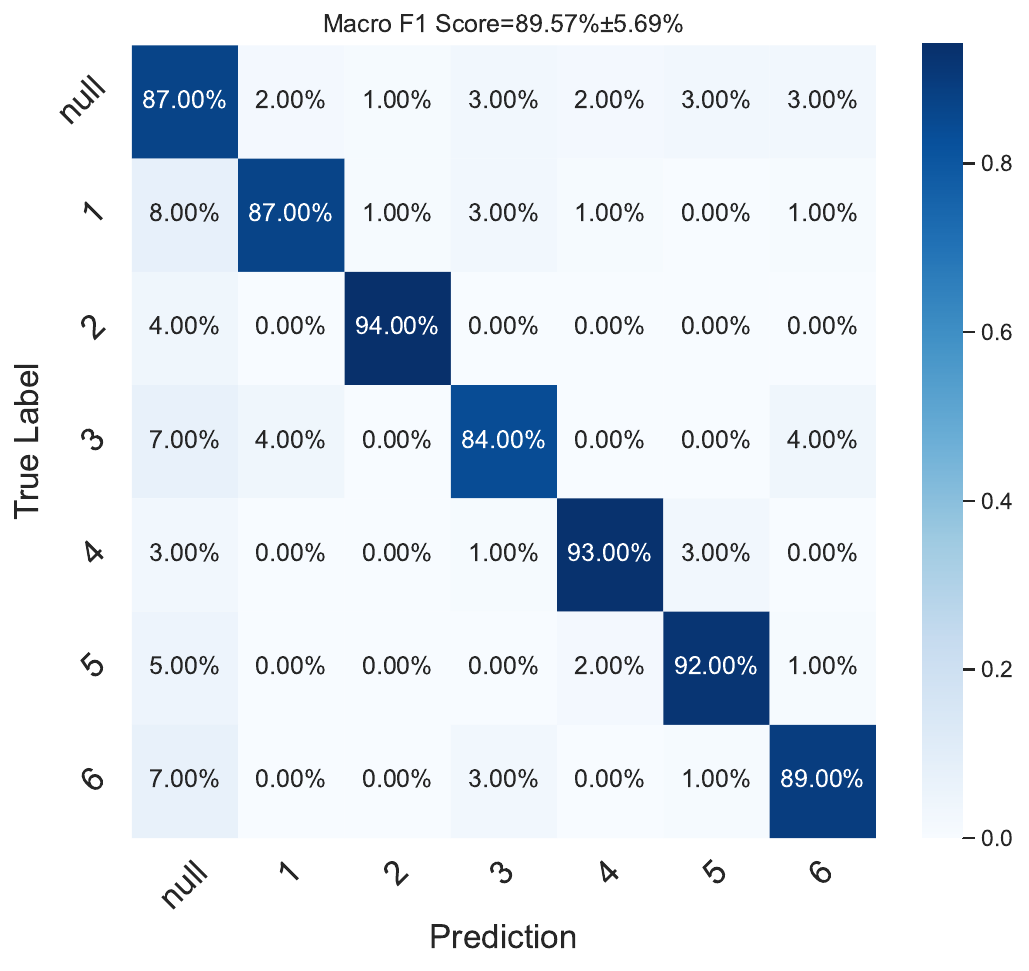}
         \caption{Bio-impedance and IMU}
         \label{fig:confusion_matrix_mpag}
     \end{subfigure}
    \caption{Joint confusion matrix for six upper body activities classification (1: Box, 2: Biceps curl, 3: Chest press, 4: Shoulder and chest press, 5: Arm hold and shoulder press, 6: Arm opener)}
    \label{fig:confusion_matrix}
\end{figure*}

\cref{fig:confusion_matrix_imp} shows the classification results with a single bio-impedance sensing modality. 
An average Macro F1 score of 75.36 \% was achieved.
The most confusing classes are \textit{Arm hold and shoulder press} and \textit{Arm opener}.
Besides, more than eleven percent of instances from the two activities, such as \textit{Arm opener} and \textit{Chest press}, are recognized as a null class. 
These two kinds of activity mainly require only the movement of the shoulder instead of bending the arm resulting in smaller bio-impedance variation compared to other activities. 
\cref{fig:confusion_matrix_imu} shows the classification result from a single IMU sensor, which achieved an average mMcro F1 score of 81.49 \%. 
The most confusing two classes are the \textit{Chest press} and \textit{Arm opener}. 
These two activities have a similar movement trajectory, where the subject presses the dumbbells away from the chest until the arms are fully extended; the difference is that the former activity requires the subject to bend the arms, and the latter requires the subject to straighten the arms, which can confuse IMU-only models as the raw IMU data was the norm of the acceleration together with they gyro's norm.
It can be observed that the average Macro F1 score of IMU is better than bio-impedance sensing in this work, which may be explained by the variation of absolute bio-impedance value across subjects and less sensitivity to shoulder movement of bio-impedance sensing in this study.

We also evaluated the performance of using both bio-impedance and IMU data as input to our encoder.
In this data-level fusion the data from two sensors was concatenated directly before input into the Backbone model. 
\cref{fig:confusion_matrix_mpag} illustrated the result of two sensing modalities fusion method. 
The average Macro F1 score was improved to 89.57 \%, which is higher than the use of a single sensing modality.
It can be observed from the confusion matrix that activities confused by the model with single modality input can be well recognized after fusing IMU and bio-impedance data; the recognition recall of these activities was improved by around 10 \%. 
The significant improvement in classification results demonstrated that the bio-impedance sensing modality could provide complementary information to the IMU sensor, achieving a better result for the fitness activity recognition task with a data-level sensor fusion method.
A result summary and model size of the Backbone model with different input modalities is shown \cref{tab:baseline_result}.
We can also see there that the concurrent use of multiple modalities with the data-level sensor fusion method can significantly improve the activity recognition performance, but also leads to an increase in model size.

\begin{table}
    \centering
    \caption{Result summary of backbone network for classifying upper body activities}
    \begin{tabular}{cccc}
    \hline
        Sensor modality & Macro F1 (\%) & Accuracy (\%) & Model Parameters\\
    \hline
     Bio-impedance    &75.36 ±12.78  & 76.19 ±11.63 & 21829\\ 
     \hline
     IMU              &81.49 ±7.24  & 81.62 ±6.63 & 21829\\
     \hline
     Bio-impedance \\ and IMU  & 89.57 ±5.69 & 89.39 ±5.82 & 23429\\
    \hline
    \end{tabular}
    
    \label{tab:baseline_result}
\end{table}

\section{Using bio-impedance sensing to train better IMU-only classification models}

We have shown that the combination of IMU and bio-impedance sensing can improve classification performance by providing crucial information that is lacking for each modality alone.
On the other hand, from the two modalities, the IMU one has many advantages. It has a better overall performance than the bio-impedance, but, even more relevant, it is widely available and of easy deployment in real-life scenarios. While it is not hard to collect data for both modalities in a laboratory environment, real deployment requires attaching the electrodes to the body parts and prolonged measurement may cause discomfort to the user. This could prohibit them from being widely used for extended periods of time.

To address this issue, we propose a constrastive-learning based approach for improving IMU representations that uses bio-impedance \emph{only at training time}. Our setup was based on the existing contrastive learning method \cite{fortes2022learning}. While in that work authors explored how extra IMU sensors only available at training time can be used to improve target IMU representations, here we extend the work to include additional different modalities at training time and propose a hyperparameter to control the knowledge transfer amount and direction. 


For training those networks, we follow a two step procedure as in \cite{fortes2022learning}, where first the correspondence between the representations
is trained based on contrastive loss. But, different from the existing work \cite{fortes2022learning}, we use a weighted contrastive loss function with a  hyperparameter $\alpha$ to control the knowledge transfer between source latent representation and target latent representation.
Let's go into detail on how this loss is computed.
If we sample  $x_{src}$ and $x_{tgt}$  windows for source and target modalities, respectively, where every  $x^{i}_{src}$ and $x^{i}_{tgt}$ co-occur
in time, we can obtain for each a representation by computing $R_{t} = Enc_{tgt}(x_{tgt})$ and $R_{s} = Enc_{src}(x_{src})$ . Then, those representations
can be translated between one another by calculating $T_{t2s} = Trl_{t2s}(R_{t})$ and $T_{s2t} = Trl_{s2t}(R_{s})$. Our
proposed weighted contrastive loss $L_{WN}$ is defined as
\begin{equation}
    \label{eq: weight_contrastive_loss}
    \begin{aligned}
    \mathbb{L}_{WN}(R_{t}, T_{s2t}, R_{s}, T_{t2s}, \alpha) = \alpha * L_{CO} (R_{t}, T_{s2t}) +  \\
                                            (1 - \alpha) * L_{CO} (R_{s}, T_{t2s})
    \end{aligned}
\end{equation}
which is a version of the one proposed in \cite{fortes2022learning} where we weight the transfer of knowledge
(from source to target and target to source). The $L_{CO}$ part of the loss is the infoNCE loss \cite{oord2018representation}
defined as  
\begin{equation}
            \label{eq: contrastive_loss_true}
				\begin{split}
					L_{CO}(X_{r}, X_{t}) =   \frac{1}{N} \sum_k \Big(
					& L_{N}(
					X_{r}^{k},
					X_{t}^{k},
					\big\{ X_{r}^{i}\big\}^{N}_{i=0}
					) \Big)
				\end{split}
\end{equation}
shows. Here $L_{N}$ is
\begin{equation}
    \label{eq: contrastive_loss_sim}
    L_{N}(x,x_t,P_x) = -log \frac{exp(sim(x_t,x))/\tau}{\sum_{x_p \epsilon P_x}exp(sim(x_p,x))/\tau}
\end{equation}
where $\tau$ is a temperature hyper-parameter configured as 0.1 in this work, the $sim(x_i,x_j)$ is defined as \cref{eq: consine_Similarity_loss} shows, 
\begin{equation}
    \label{eq: consine_Similarity_loss}
    sim(x_i,x_j) = \frac{x^{T}_{i}\cdot x_j}{\left\| x_i \right\|\cdot \left\| x_j \right\|}
\end{equation}

After $80$ epochs of training the correspondence between representations, we are ready for the second step of our method,
which involves fine-tuning the learned representation  for classification. 
We obtain predictions using a single classifier network $C$. For the target domain, we can obtain those by applying the classifier on the target's representation using
\begin{equation}
    \label{eq: pred_target}
    P_t = C(Enc_{tgt}(x_{tgt}))
\end{equation}
For the source domain, we will also use the same $C$ network, but this time it will receive translated points to the target's representation as in \cref{eq: pred_source}
\begin{equation}
    \label{eq: pred_source}
    P_s = C(Trl_{s2t}(Enc_{src}(x_{src})))
\end{equation}
The classification function $\mathbb{L}_c$ was defined as
\begin{equation}
    \begin{aligned}
    \label{eq: classification_loss}
    \mathbb{L}_c(P_s, P_t, y) = CE(P_s, y) + CE(P_t, y) \\
    \end{aligned}
\end{equation}
Where $y$ is the labels, $C$ is our classifier network and $CE$ is the cross-entropy-loss function. 
This second step runs for  $20$ epochs and the full training procedure of CL model can be found in\cref{al:training procedure}.
After training we can evaluate our model on the target domain using $C(Enc_{tgt}(x_{tgt}))$.

\begin{figure}[t]
\centering
\includegraphics[width=1.0\linewidth]{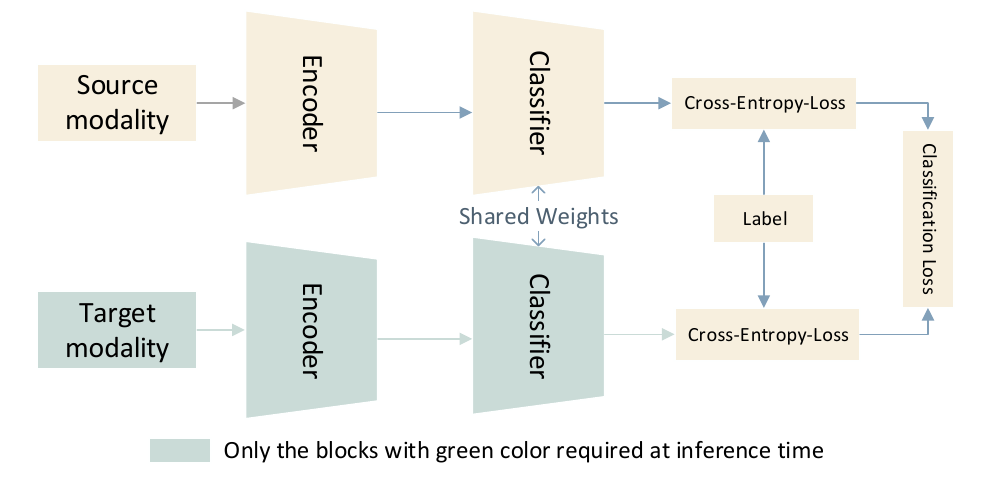}
\caption{ Cross-modality implementation with Shared Representation (SR) }
\label{fig:SharedRepresentationModel}
\end{figure}

To evaluate the proposed contrastive learning approach, two types of source modalities were designed: just bio-impedance and the combination of bio-impedance and IMU.
In order to investigate the effect of having this shared classifier between sources, we also trained using only classification loss 
a shared representation model as in \cref{fig:SharedRepresentationModel}. The shared representation model has the same architecture as the contrastive model at the inference time, but has no translators. Thus, in this case for data $x_{src}$ and $ x_{tgt}$
\begin{equation}
    \begin{aligned}
    \label{eq: classification_loss_shared}
    \mathbb{L}_{sr}(x_{src}, x_{tgt}, y) = CE( C(Enc_{src}(x_{src})) , y) \\
    	 + CE( C(Enc_{tgt}(x_{tgt})) , y) 
    \end{aligned}
\end{equation}
In this case it is trained for $100$ epochs with early stopping and $30$ patience.

\begin{figure*}[t]
\footnotesize
\centering
\includegraphics[width=1.0\linewidth]{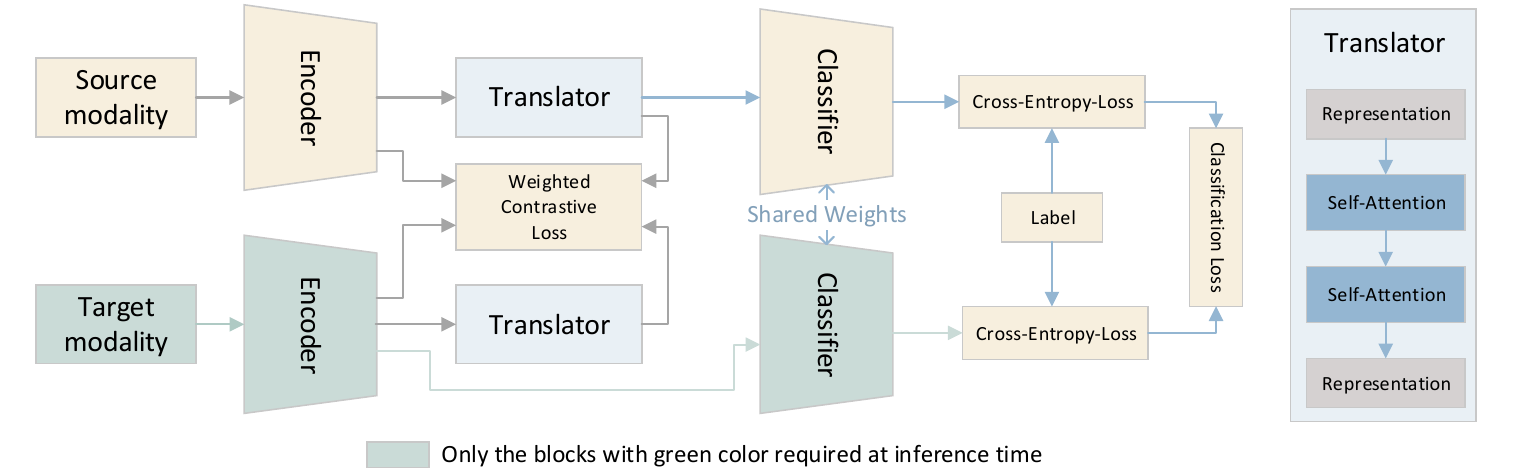}
\caption{Proposed cross-modality implementation with contrastive Loss}
\label{fig: architecture impedance net}
\end{figure*}

\begin{algorithm}
\caption{Training procedure of our contrastive approach. We used the default values of  
    $m=1024$,
    $\alpha=1.0$,
    $nEp_{constr}=80$,
    $nEp_{classif}=20$}
\begin{algorithmic}
\label{al:training procedure}
\REQUIRE Batch size $m$,
    Adam hyperparameter $\eta$, hyperparameters $alpha$,  $NEP_{constr}$, $NEP_{classif}$

\STATE \textbf{Input:} $X_{src}$, $X_{tgt}$, $Y$ representing labelled synchronised sources.

\FOR{number of batches in $NEP_{constr}$ epochs}
    \STATE Sample a batch $(x_{s},x_{t})$ with $m$ examples from  $X_{src}$, $X_{tgt}$ corresponding to the data at the same windows.
    \STATE $R_s \gets Enc_{src}(x_{s})$
    \STATE $R_t \gets Enc_{tgt}(x_{t})$
    \STATE $T_{s2t} \gets Trl_{s2t}(R_s)$
    \STATE $T_{t2s} \gets Trl_{t2s}(R_t)$
    \STATE $\theta \gets \theta - \eta \nabla_{\theta}  \mathbb{L}_{WN}(R_{t}, T_{s2t}, R_{s}, T_{t2s}, \alpha)$
    \hfill$\triangleright$\cref{eq: weight_contrastive_loss}
\ENDFOR

\FOR{number of batches in $NEP_{classif}$ epochs}
    \STATE Sample a batch $(x_{s},x_{t},y_{l})$ with $m$ examples from  $X_{src}$, $X_{tgt}$, $Y$ corresponding to the data at the same windows.
    \STATE $P_s \gets C(Trl_{s2t}(Enc_{src}(x_{s})))$
    \STATE $P_t \gets C(Enc_{tgt}(x_{t}))$
    \STATE $\theta \gets \theta - \eta \nabla_{\theta} \mathbb{L}_{c}(P_s, P_t, y_{l})$
    \hfill$\triangleright$\cref{eq: classification_loss}
\ENDFOR
\end{algorithmic}
\end{algorithm}

\begin{figure}[t]
\centering
\includegraphics[width=1.0\linewidth]{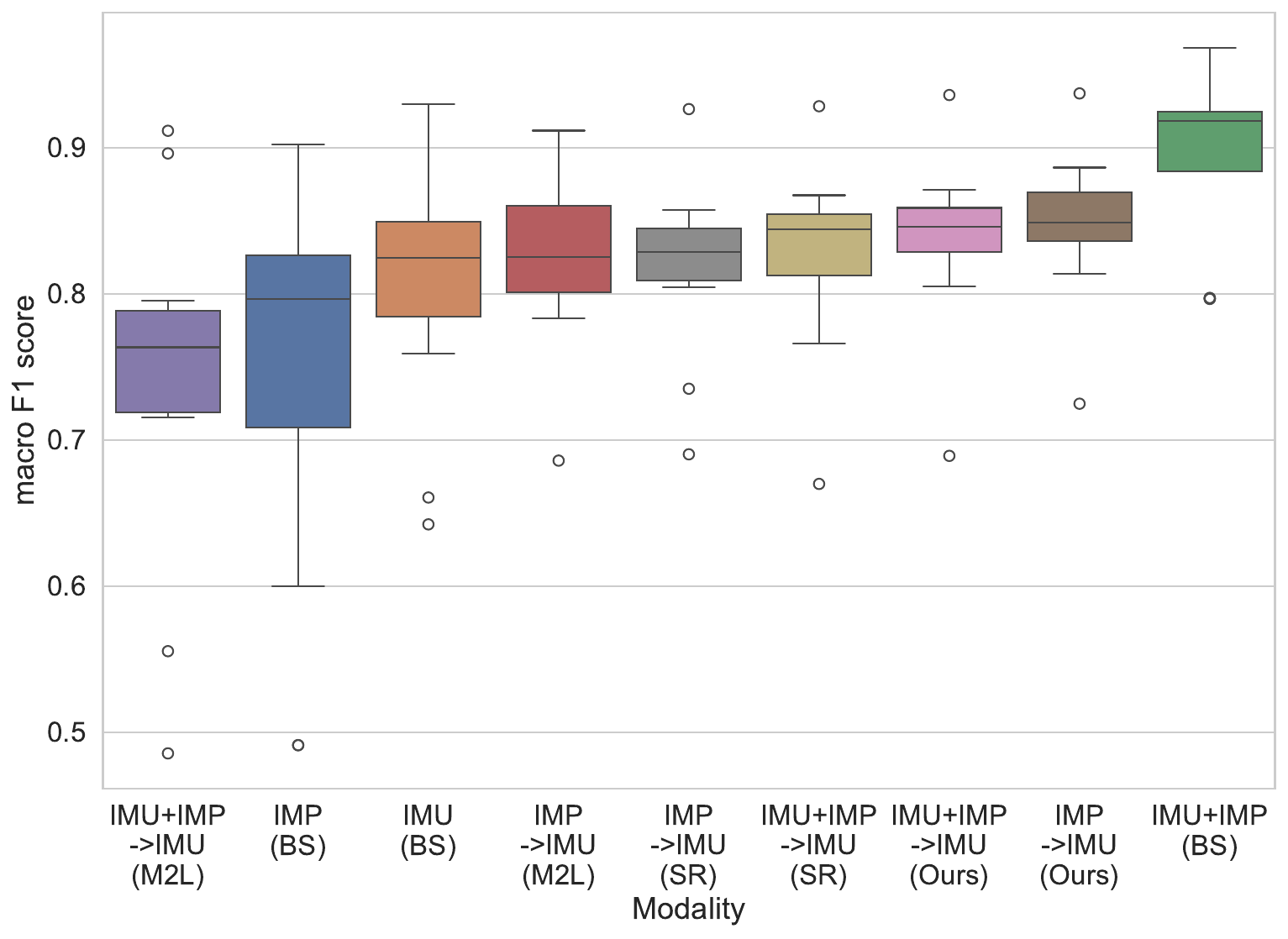}
\caption{Boxplots of the result for upper body activities with different models (\textbf{IMP}: Bio-impedance. \textbf{BS}: baseline from Backbone model. \textbf{M2L}: model from existing work \cite{yang2022more}. \textbf{SR}: shared representation model. \textbf{Ours}: Contrastive learning (CL) based model. In the M2L, SR, and our models, the left of the arrow is the source modality and the right side is the target modality}
\label{fig:box_plot_result}
\end{figure}

\begin{figure}[t]
\centering
\includegraphics[width=1.0\linewidth]{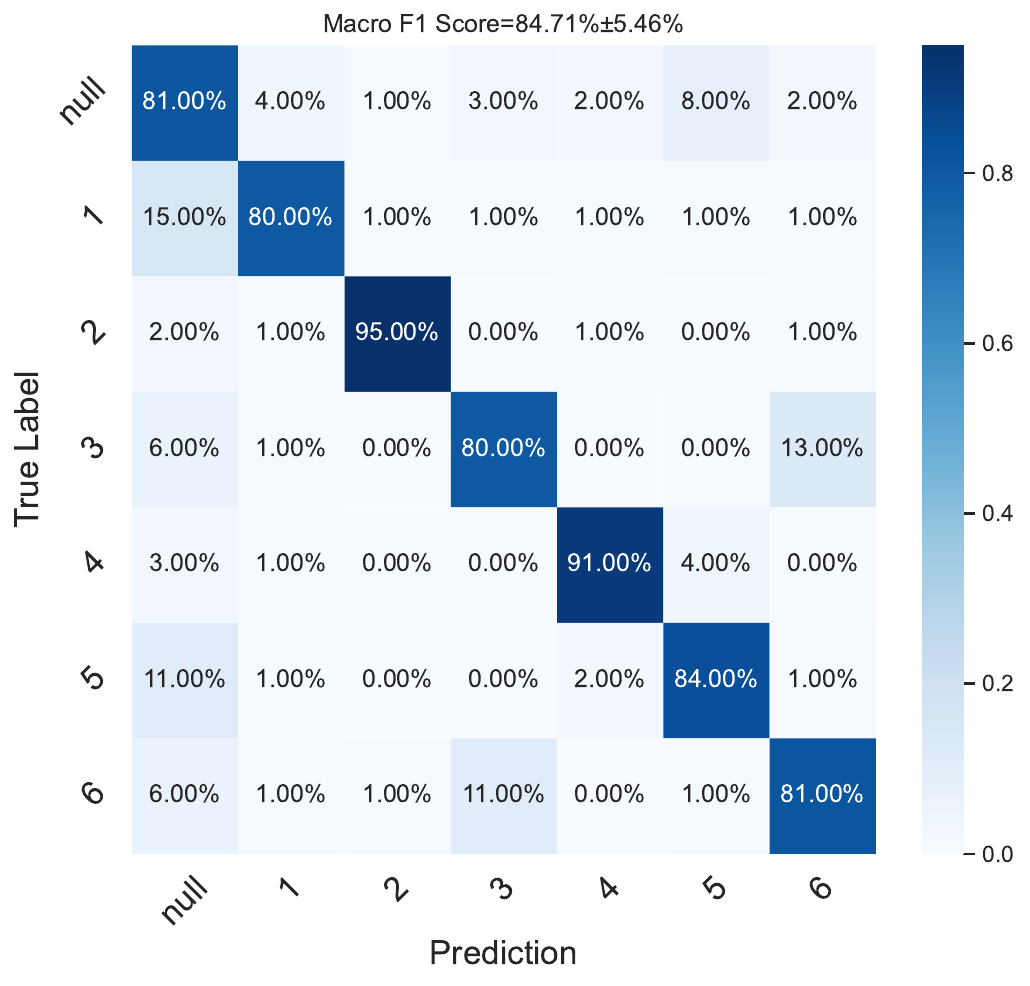}
\caption{Joint confusion matrix of our model based on contrastive learning with the input of single IMU at inference(1: Box, 2: Biceps curl, 3: Chest press, 4: Shoulder and chest press, 5: Arm hold and shoulder press, 6: Arm opener)}
\label{fig:matrix_result_CL model}
\end{figure}

\begin{table*}
\footnotesize
    \centering
    \caption{Result summary of upper body activity recognition}
    \begin{tabular}{cccccc}
        \hline
         Model&  Target modality&  Source modality&  Macro F1 Score(\%) &Weight ($\alpha$)& Improvement (\%)\\
        \hline
        
        
        \hline
        
        Backbone & IMU & - & 81.49 ± 7.24 & - & - \\
        \hline

        \multirow{ 2}{*}{M2L \cite{yang2022more}}  & IMU & Bio-impedance &   80.53\% ± 7.43\%&  -& -0.96\\
                               & IMU & Bio-impedance and IMU & 79.80\% ± 6.90\%& -& -1.69\\
        \hline                    
        \multirow{ 2}{*}{SR}  & IMU & Bio-impedance &   81.74 ± 6.51&  -& 0.25\\
                                        & IMU & Bio-impedance and IMU & 82.57 ± 6.88& -& 1.08\\

        \hline
        \multirow{ 2}{*}{Ours}  & IMU & Bio-impedance &   84.71 ± 5.46&  1.0& \textbf{3.22}\\
                                        & IMU & Bio-impedance and IMU & 83.64 ± 6.23& 0.1& 2.15\\
         \hline

        \hline
        Backbone & Bio-impedance & - & 75.36 ± 12.78 &  & \\

        \hline
        \multirow{ 2}{*}{M2L \cite{yang2022more}} & Bio-impedance & IMU &   65.76\% ± 14.66\%&  -& -9.60\\
                                                & Bio-impedance & IMU and Bio-impedance & 66.40\% ± 13.64\%& -& -8,96\\
                                                
        \hline
        \multirow{ 2}{*}{SR} & Bio-impedance & IMU &   74.22 ± 13.91&  -& -1.14\\
                                                & Bio-impedance & IMU and Bio-impedance & 75.05 ± 13.71& -& -0.31\\
        \hline
        \multirow{ 2}{*}{Ours}          & Bio-impedance & IMU &   77.76 ± 12.71&  0.3& \textbf{2.40}\\
                                                & Bio-impedance & IMU and Bio-impedance & 75.74 ± 13.00& 0.2& 0.38\\
        \hline
        \hline
        Sensor Fusion & IMU and Bio-impedance & - & 89.57 ± 5.69  &  &\\
        \hline
    \end{tabular}
    
    \label{tab:result_summary}
\end{table*}

Since the hyperparameter weight $\alpha$ can not be decided before training directly, a grid search method with a search space from 0 to 1.0 with a step length of 0.1 was used to select the best weight value during training our contrastive learning model.

\subsection{Results}

\cref{fig:box_plot_result} illustrates the distribution of classification performance for models with different source modality combinations and training procedures; it can be observed that both the shared representation model and our contrastive training approach achieved a better performance than the models trained with only one sensor modality, which shows that learning the temporal correspondence between our modalities can positively guide the representation in our single-modality models.
The standard deviation of the test result is also decreased when the models are trained with two modalities.
In addition, our proposed model trained by weighted contrastive learning loss showed superior performance compared to the shared representation model only using classification loss.
When the target modality is IMU, the best average macro F1 score was improved by 3.22 \% by our contrastive learning approach with $\alpha=1.0$ as shown in \cref{tab:result_summary}, which means that the contrastive learning loss is computed using only $R_{t}$ and $T_{s2t}$ according to \cref{eq: weight_contrastive_loss}.
It is worth mentioning that a single bio-impedance modality demonstrated a better performance than using both bio-impedance and IMU as source, which achieved the best performance with an average Macro F1 score with 2.15 \% when the weight $\alpha$ was configured as 0.1. 

In \cref{tab:result_summary}, we have also compared our method to M2L\cite{yang2022more}. In this case, we used the same hyper-parameters as in the original work, but kept the architecture of models as in \cref{fig:SharedRepresentationModel} for fairness, with the important difference that in this case the classifiers do not share weights, as is the case in their method. From the results, it is clear that our contrastive learning approach outperforms it, as it did not improve on the baseline with our studied modality.

\cref{fig:matrix_result_CL model} shows the joint confusion matrix for the contrastive learning model with the source modality of a single bio-impedance while the weight $\alpha$ was 1.0. 
The recall of each class has been improved from 1\% to 5\% by our approach compared to the result of the model trained with a single IMU as shown in \cref{fig:confusion_matrix_imu}.
The activities with the most improved recall of 5 \% are \textit{Arm hold+Shoulder Press} and \textit{Box}. 
The class \textit{Biceps Curl} with the highest recognition recall score of 95 \% was achieved by our model, which is even higher than the direct sensor fusion method with a recall of 94 \% as \cref{fig:confusion_matrix_mpag} shown, where the bio-impedance and IMU data are required at inference time.


As our proposed model has demonstrated promising performance in using bio-impedance sensing to train better IMU-only classification, we further investigated whether the same approach can train better bio-impedance-only classification using IMU data, where the bio-impedance was regarded as the target sensor while the IMU or IMU and Bio-impedance sensor were the source target.
As a result in \cref{tab:result_summary} shows, the average Macro F1 score with a single bio-impedance modality as the modality at inference time was improved by 2.24 \% with the help of IMU sensor as source modality during training compared to the backbone classification model.
However, there is a performance degradation of the Macro F1 score for this task by the SR model and M2L model.
Overall, the proposed model based on the weighted contrastive loss can be used to train a better IMU-only classification with the help of the source sensor modality, which is not required at the inference time.
The performance of our model is better than the model trained only by the target sensor modality, but cannot achieve the same performance as the direct sensor fusion model, which requires all modalities to be available at both training and inference time. This is because the temporal correspondence between the modalities can only guide the target modality to a better representation, but clearly there is information missing as the source modality is not present at inference time.

\begin{figure}[t]
\centering
\includegraphics[width=1.0\linewidth]{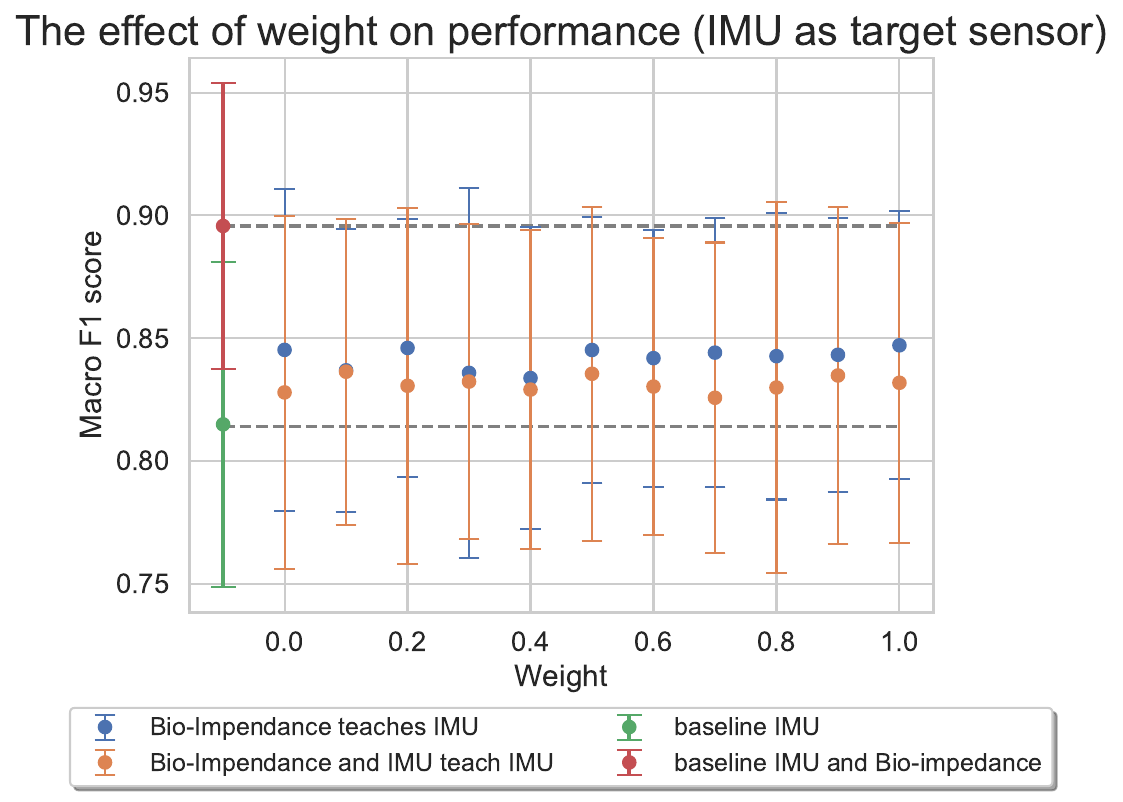}
\caption{The effect of weight on performance when  IMU is the target sensor (upper body activity recognition)}
\label{fig:weight_vs_performance_imu}
\end{figure}

\begin{figure}[t]
\centering
\includegraphics[width=1.0\linewidth]{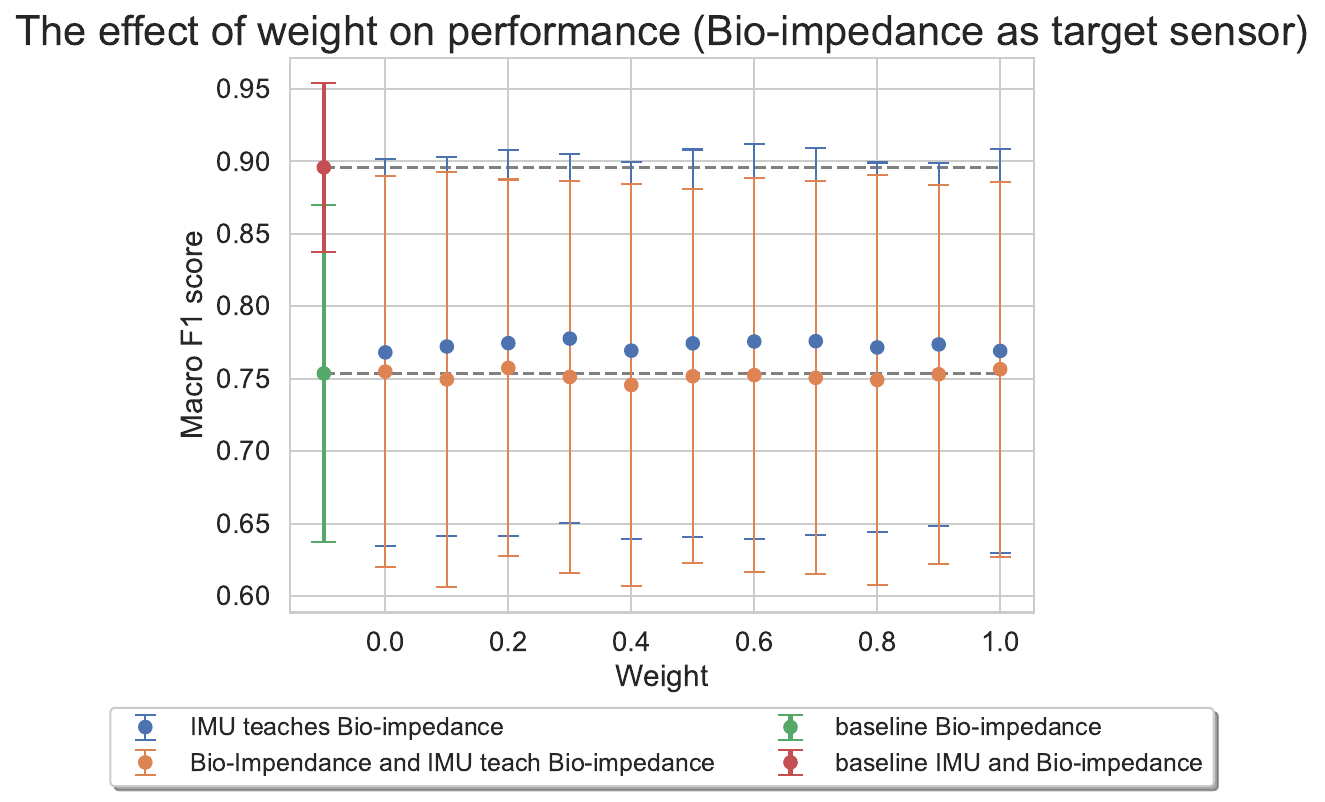}
\caption{The effect of weight on performance when the bio-impedance is the target sensor (upper body activity recognition)}
\label{fig:weight_vs_performance_imp}
\end{figure}

Since the proposed hyperparameter weight $\alpha$ can affect the latent representation translation, the performance of our contrastive learning model is closely related to this parameter.
In order to study the effect of the weight on classification result, eleven weight value from 0 to 1 was tested,
\cref{fig:weight_vs_performance_imu} shows the effect of weight on the performance of classification task when the IMU as the target sensor modality. It can be observed that the average Macro F1 score is changed with different weight values, however, there is not a clear relationship between the weight value and the performance.
The same result can be observed in the contrastive approach with the target source of bio-impedance as shown in \cref{fig:weight_vs_performance_imp}.
The optimal weight value among different target and source modalities is also different.

\section{Extended Study on Lower Body Fitness Activity Recognition}

\begin{figure*}
    \centering
    \includegraphics[width=1.0\linewidth]{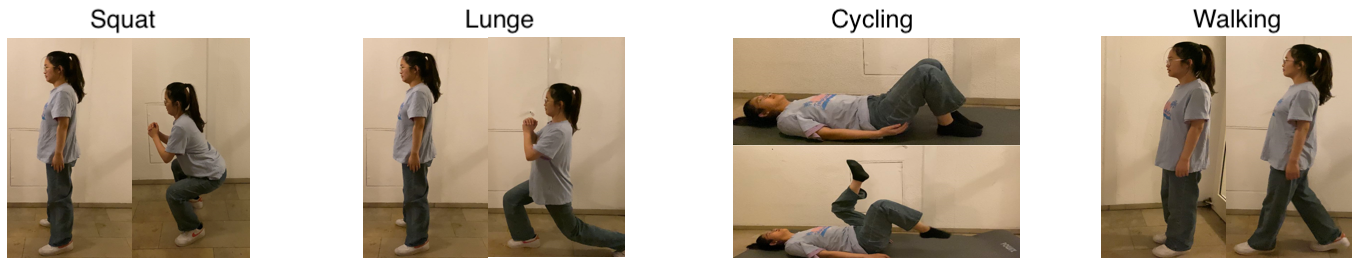}
    \caption{The four activities used for our lower body fitness activity recognition scenario. Two electrodes were attached to the two calves separately, by which the knee can be connected into the body circuit loop between the two electrodes, while the single IMU was only attached to the left calf.   }
    \label{fig:legActs}
\end{figure*}

To demonstrate the generalizability of iMove in fitness activity recognition, an extended experiment on lower body fitness activity was conducted in this work.
The experiment protocol is similar to the experiment on the upper body fitness activity except for the sensor location.
In the lower body fitness activity experiment, two electrodes were attached to the two calves separately, by which the knee can be connected into the body circuit loop between the two electrodes as the knee bending activities play an important role in lower body activity, like leg fitness activity. 
The single IMU was only attached to the left calf. 
Five types of leg activity were performed by ten subjects on five different days. The activities selected can be seen in \cref{fig:legActs} and are squat, lunge, cycling, and walking. As we can see in the Figure the cycling activity was performed with the subject lying down on an exercise mat. The other activities performed during the experiment such as standing still and activity transitions are regarded as null class.

\begin{figure*}[!t]
\footnotesize
\centering
     \begin{subfigure}[b]{0.33\textwidth}
         \centering
         \includegraphics[width=\textwidth]{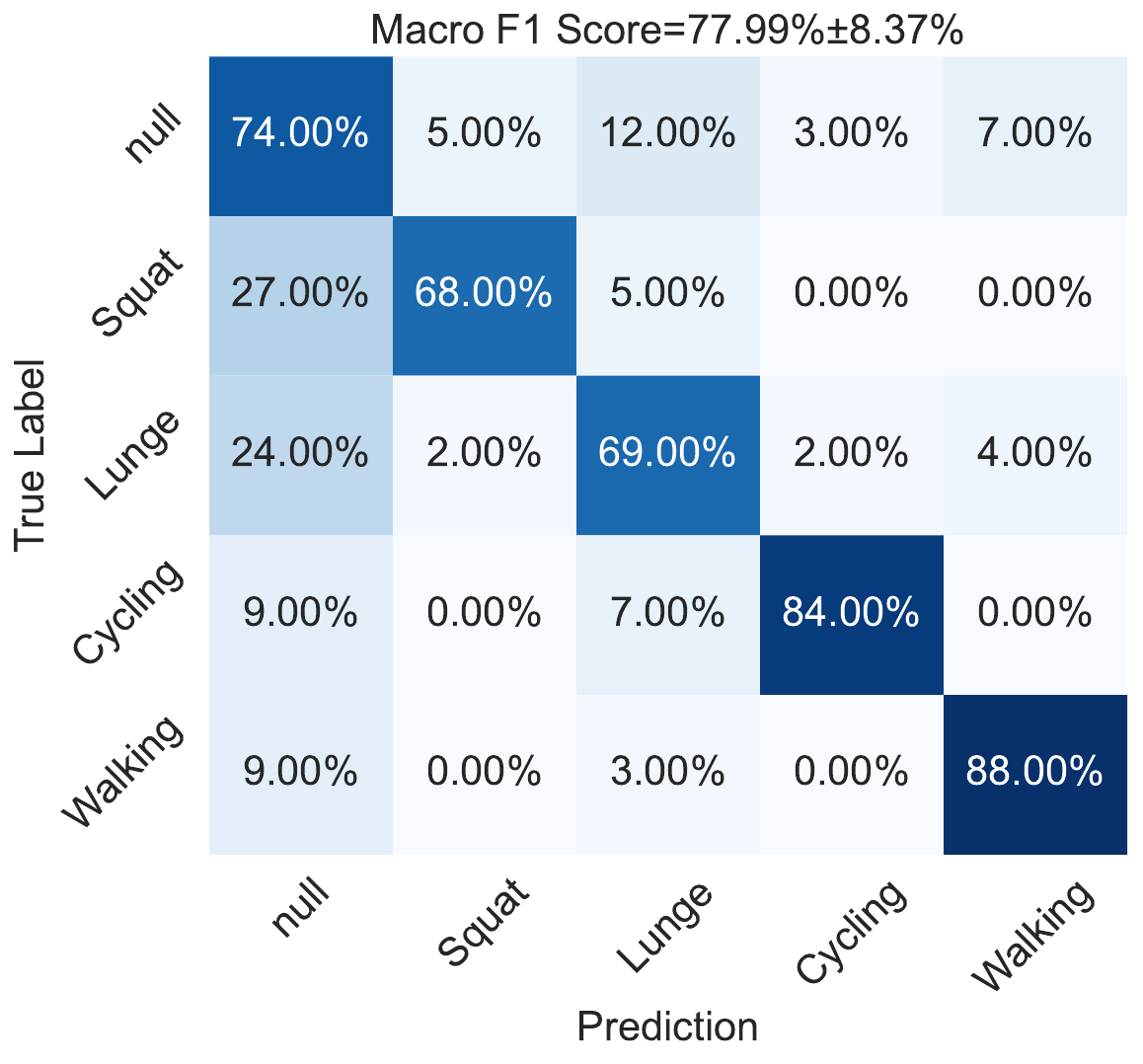}
         \caption{Single IMU}
         \label{fig:confusion_matrix_leg_ag}
     \end{subfigure}
     \hfill
     \begin{subfigure}[b]{0.33\textwidth}
         \centering
         \includegraphics[width=\textwidth]{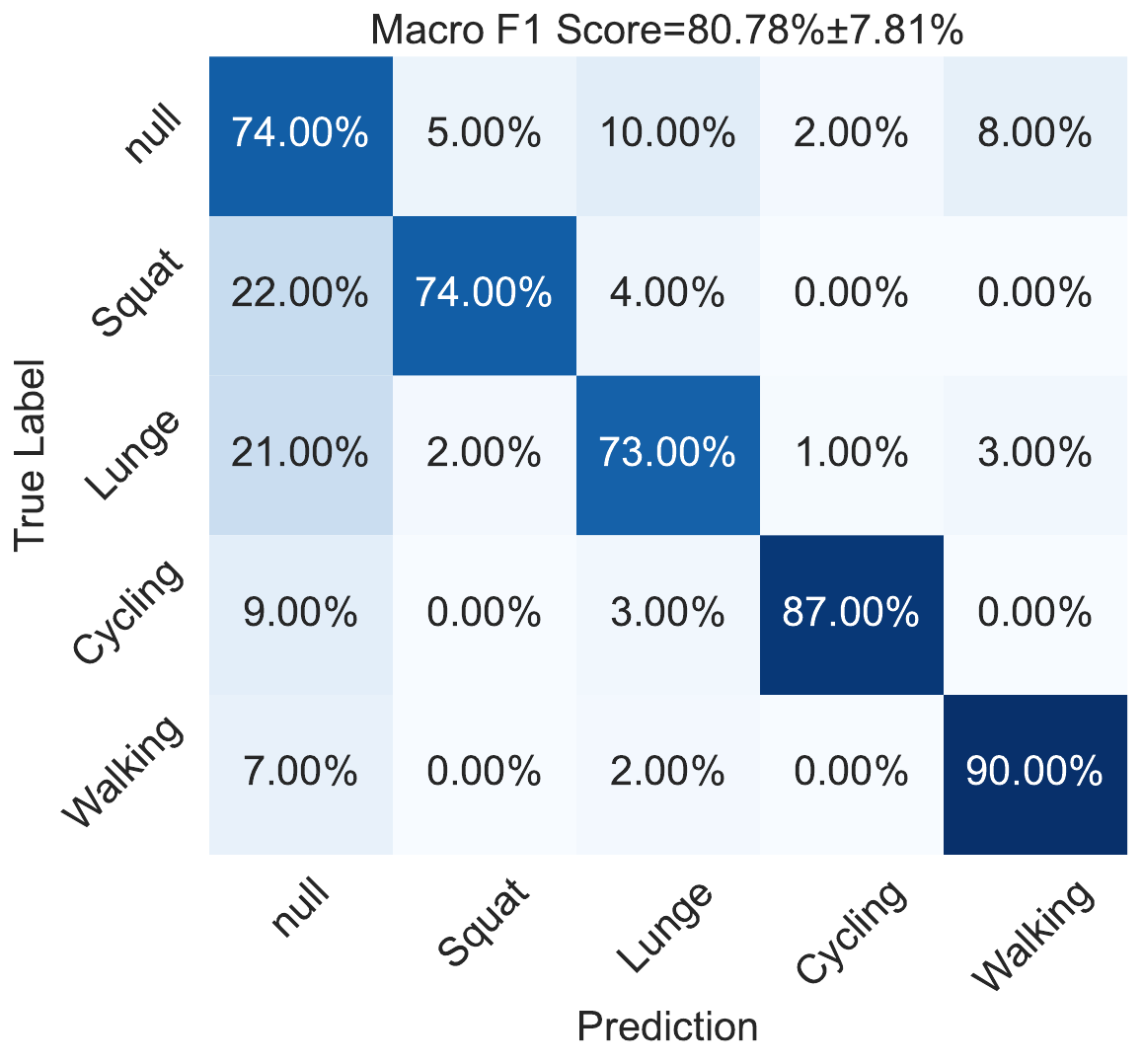}
         \caption{Bio-impedance teaching IMU (\textbf{Ours})}
         \label{fig:confusion_matrix_leg_con}
     \end{subfigure}
     \hfill
     \begin{subfigure}[b]{0.33\textwidth}
         \centering
         \includegraphics[width=\textwidth]{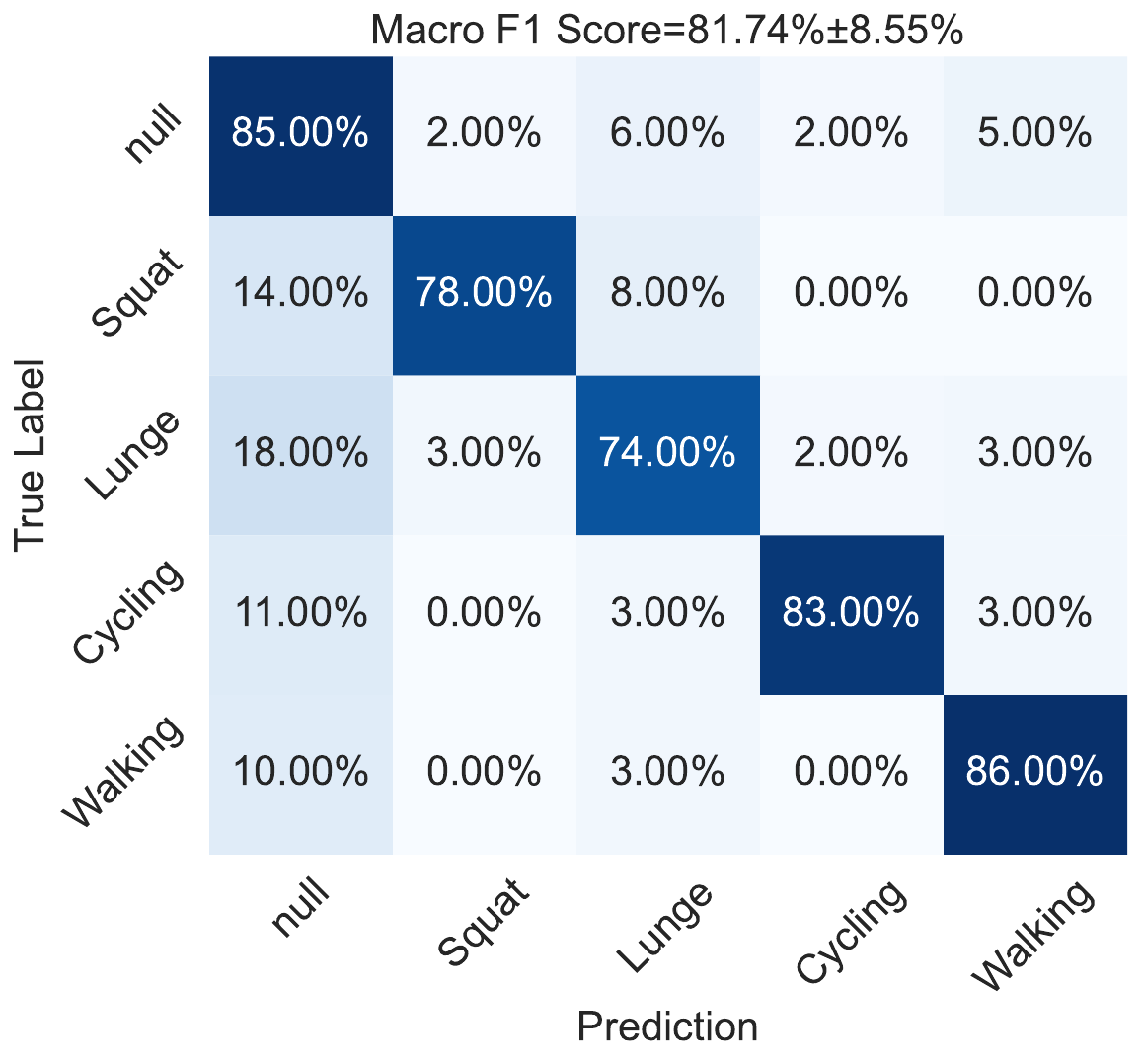}
         \caption{Bio-impedance and IMU fusion directly}
         \label{fig:confusion_matrix_leg_mpag}
     \end{subfigure}
    \caption{Joint confusion matrix for lower body fitness activity classification}
    \label{fig:confusion_matrix_leg}
\end{figure*}

\cref{fig:confusion_matrix_leg} shows the lower body fitness activity classification result.
It can be observed that the best classification result was achieved by the direct sensor fusion method with an average Macro F1 score of 81.74 \%, while the model with a single IMU as input achieved an average F1 score of 77.99 \%.
With our proposed method using Bio-impedance data to guide IMU during the model training, the classification result has been improved by around 3\% and is very close to the direct sensor fusion method as \cref{fig:confusion_matrix_leg_con} shows.
The most confusing classes are the null class and other defined activities, especially 
for the slow activities like squat and lunge, which can not be well recognized by single IMU as \cref{fig:confusion_matrix_leg_con} shows compared to walking and cycling, because the IMU sensor is on the calf while the major movement of squat is from knee blending leading to little movement in the calves.
However, the knee blending can cause bio-impedance variation between two electrodes, which provides additional information to IMU, thus, the activity classification results were improved when adding the bio-impedance information to the model as \cref{fig:confusion_matrix_leg_con} and \cref{fig:confusion_matrix_leg_mpag} show.

\begin{figure}[!t]
\centering
\includegraphics[width=1.0\linewidth]{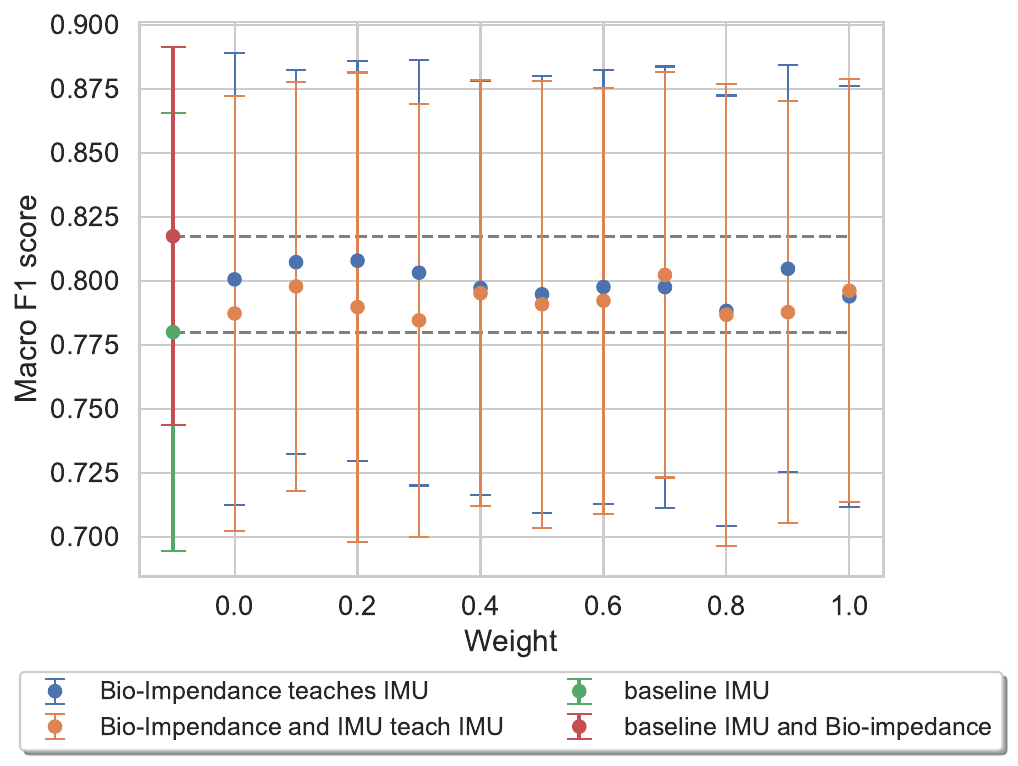}
\caption{The effect of weight on performance when  IMU is the target sensor in lower body fitness activity classification}
\label{fig:weight_vs_performance_imu_leg}
\end{figure}

\begin{figure}[!t]
\centering
\includegraphics[width=1.0\linewidth]{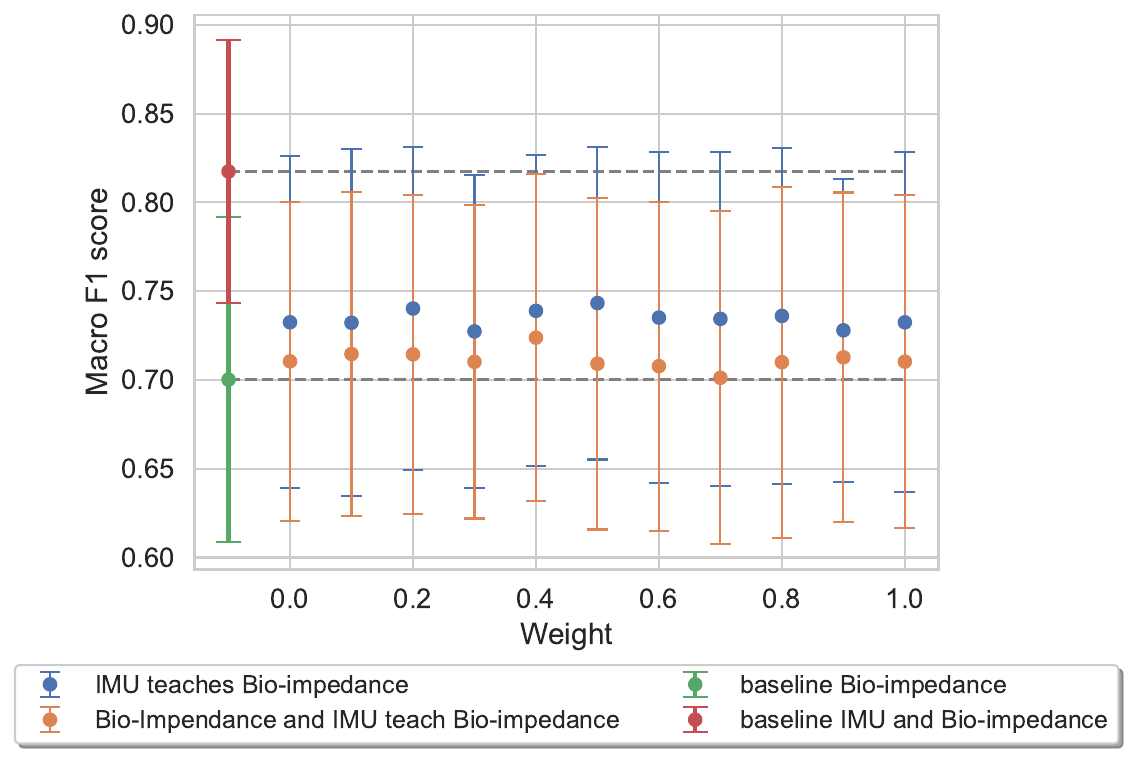}
\caption{The effect of weight on performance when  bio-impedance is the target sensor in lower body fitness activity classification}
\label{fig:weight_vs_performance_imp_leg}
\end{figure}

\cref{tab:result_summary_leg} shows the result summary of the experiment on the lower body fitness activity classification. 
The result is similar to the experiment on upper body fitness activity classification, our proposed method helps a single IMU-based inference model improve the lower body fitness activity classification Macro F1 score by 2.79 \%.
In addition, the performance of a single bio-impedance-based inference model was also improved by 4.31 \% with the use of our proposed solution.
The results outperform both the M2L model and the shared representation model.
\cref{fig:weight_vs_performance_imp_leg} and \cref{fig:weight_vs_performance_imu_leg} show the effect of weight on the performance of the leg fitness activity classification task when the IMU and Bio-impedance as the target sensor modality, correspondingly. 
Similar to the upper body fitness activity classification, the optimal weight values change depending on target and source modalities. 
In most cases, the weight with the value of 0.5, which is the same as the existing work Learning from the Best \cite{fortes2022learning}, improves results, but does not achieve the best ones. 
Thus, our proposed weighting factor plays an important role in activity classification tasks.
In future work,  we will investigate how weight selection can be done end to end in the training procedure instead of remaining a hyper-parameter.

\begin{table*}
\footnotesize
    \centering
    \caption{Result summary of lower body fitness activity recognition}
    \begin{tabular}{cccccc}
        \hline
         Model&  Target modality&  Source modality&  Macro F1 Score(\%) &Weight ($\alpha$)& Improvement (\%)\\
        \hline
        
        
        \hline
        
        Backbone & IMU & - & 77.99 ± 8.37 & - & - \\
        \hline

        \multirow{ 2}{*}{M2L \cite{yang2022more}}  & IMU & Bio-impedance &   77.13 ± 7.87 &  -& -0.86\\
                               & IMU & Bio-impedance and IMU & 77.30 ± 8.02& - & -0.69\\
        \hline                    
        \multirow{ 2}{*}{SR}  & IMU & Bio-impedance &   77.84 ±8.69 &  -&  -0.15\\
                                        & IMU & Bio-impedance and IMU & 78.53 ±7.61& -&  0.54\\

        \hline
        \multirow{ 2}{*}{Ours}  & IMU & Bio-impedance &   80.78 ±7.81&  0.2& \textbf{2.79}\\
                                        & IMU & Bio-impedance and IMU & 79.51 ±8.31 & 0.4& 1.52\\
        \hline
        \hline
        Backbone & Bio-impedance & - & 70.02 ± 10.09 &  & \\

        \hline
        \multirow{ 2}{*}{M2L \cite{yang2022more}} & Bio-impedance & IMU &   70.08 ± 8.96&  -& 0.06\\
                                                & Bio-impedance & IMU and Bio-impedance & 71.29 ± 9.23& & 1.27\\
                                                
        \hline
        \multirow{ 2}{*}{SR} & Bio-impedance & IMU &   70.71 ± 10.72&  -& 0.69\\
                                                & Bio-impedance & IMU and Bio-impedance & 71.60 ± 10.44& -& 1.58\\
        \hline
        \multirow{ 2}{*}{Ours}          & Bio-impedance & IMU &   74.33 ± 8.81&  0.5& \textbf{4.31}\\
                                                & Bio-impedance & IMU and Bio-impedance & 72.38 ± 9.23& 0.4& 2.36\\
        \hline
        \hline
        Sensor Fusion & IMU and Bio-impedance & - & 81.74 ± 8.55  &  &\\
        \hline
    \end{tabular}
    
    \label{tab:result_summary_leg}
\end{table*}

\section{Conclusion}
Our research has successfully showcased the potential of bio-impedance in enhancing the accuracy of fitness activity recognition when combined with inertial measurement units (IMUs) through sensor fusion and contrastive learning techniques. Our contrastive training framework, leveraging shared latent representations from multiple modalities, enabled us to harness the power of bio-impedance to enhance a mono-modality IMU model, achieving an impressive 84.71 \% Macro F1 score. Importantly, this incorporation of bio-impedance was solely necessary during the training phase. We have evaluated our method first on six upper body fitness activities and then on four lower body ones obtaining similar results. This result highlights the importance of training rich representations for HAR, even if those are only available at training time due to deployment constraints.

Regarding bio-impedance as a modality for HAR, we have demonstrated the effectiveness of single-channel bio-impedance sensing between wrists for the recognition of fitness activities, achieving an average Macro F1 score of 75.36 \% for six upper body ones. While this performance was not as good as that of a single IMU sensor on one wrist (81.49 \%), we showed how bio-impedance can improve IMU HAR through sensor fusion, that is, by employing a data-level sensor fusion approach which provided in this case a substantial improvement with an activity classification Macro F1 of 89.57 \%. We have reached similar results for lower body activities, with once again the best results being those using sensor-fusion.

Our findings underscore the potential of sensor fusion and contrastive learning as valuable tools for advancing fitness activity recognition, with bio-impedance playing a pivotal role in augmenting the capabilities of IMU-based systems. This research opens avenues for further exploration and application of multi-modal sensing in the realm of fitness tracking and beyond.
\section*{Acknowledgment}
The research reported in this paper was supported by the BMBF (German Federal Ministry of Education and Research) in the project VidGenSense (01IW21003) and SocialWear (01IW20002).

\bibliographystyle{IEEEtran}
\bibliography{ref}

\end{document}